\documentclass{emulateapj}

\usepackage{url}
\usepackage{epstopdf}
\usepackage{multirow}

\usepackage{amsmath}
\usepackage{natbib}
\usepackage{color}
\usepackage{booktabs}
\usepackage[utf8]{inputenc}
\usepackage{tabularx}
\usepackage[colorlinks, citecolor=blue, linkcolor=blue, menucolor=blue, urlcolor=blue]{hyperref}

\slugcomment{}

\shortauthors{Park et al.}

\begin{document}

\title{Faraday rotation in the jet of M87 inside the Bondi radius: indication of winds from hot accretion flows confining the relativistic jet}

\author{Jongho Park\altaffilmark{1}, Kazuhiro Hada\altaffilmark{2,3}, Motoki Kino\altaffilmark{4,5}, Masanori Nakamura\altaffilmark{6}, Hyunwook Ro\altaffilmark{7}, and Sascha Trippe\altaffilmark{1}}
\affil{$^1$Department of Physics and Astronomy, Seoul National University, Gwanak-gu, Seoul 08826, South Korea; jhpark@astro.snu.ac.kr\\
$^2$Mizusawa VLBI Observatory, National Astronomical Observatory of Japan, 2-21-1 Osawa, Mitaka, Tokyo 181-8588, Japan\\
  $^3$Department of Astronomical Science, The Graduate University for Advanced Studies (SOKENDAI), 2-21-1 Osawa, Mitaka, Tokyo 181-8588, Japan\\
  $^4$National Astronomical Observatory of Japan, 2-21-1 Osawa, Mitaka, Tokyo, 181-8588, Japan\\
$^5$Kogakuin University of Technology \& Engineering, Academic Support Center, 2665-1 Nakano, Hachioji, Tokyo 192-0015, Japan\\
$^6$ Institute of Astronomy and Astrophysics, Academia Sinica, P.O. Box 23-141, Taipei 10617, Taiwan\\
 $^7$Department of Astronomy, Yonsei University, 134 Shinchondong, Seodaemungu, Seoul 120-749, Republic of Korea}
\received{...}
\accepted{...}

\begin{abstract}
We study Faraday rotation in the jet of M87 inside the Bondi radius using eight Very Long Baseline Array data sets, one at 8 GHz, four at 5 GHz, and three at 2 GHz. We obtain Faraday rotation measures (RMs) measured across the bandwidth of each data set. We find that the magnitude of RM systematically decreases with increasing distance from the black hole from 5,000 to 200,000 Schwarzschild radii. The data, showing predominantly negative RM sign without significant difference of the RMs on the northern and southern jet edges, suggest that the spatial extent of the Faraday screen is much larger than the jet. We apply models of hot accretion flows, thought to be prevalent in active galactic nuclei having relatively low luminosity such as M87, and find that the decrease of RM is described well by a gas density profile $\rho \propto r^{-1}$. This behavior matches the theoretically expected signature of substantial winds, nonrelativistic un-collimated gas outflows from hot accretion flows, which is consistent with the results of various numerical simulations. The pressure profile inferred from the density profile is flat enough to collimate the jet, which can result in gradual acceleration of the jet in a magneto-hydrodynamical process. This picture is in good agreement with the observed gradual collimation and acceleration of the M87 jet inside the Bondi radius. The dominance of negative RMs suggests that jet and wind axis are misaligned such that the jet emission exposes only one side of the toroidal magnetic fields permeating the winds.
\end{abstract}



\section{Introduction}

Active galactic nuclei (AGNs) are powered by accretion of gas onto supermassive black holes at the centres of galaxies. It is now widely believed that there are two distinct modes of black hole accretion: cold and hot. A cold accretion flow forms an optically thick but geometrically thin disk, radiating thermal blackbody emission with the gas temperature in the range of $10^4-10^7$ K (\citealt{SS1973}, see e.g., \citealt{Netzer2013} for a review). On the other hand, hot accretion flows are thought to be optically thin but geometrically thick with a large portion of the gravitational binding energy of the accreted gas advected into the black hole (e.g., \citealt{Ichimaru1977, NY1994}, see e.g., \citealt{YN2014} for a review). The most critical factor in determining the accretion mode is the mass accretion rate ($\dot{M}$) relative to the Eddington rate ($\dot{M}_{\rm Edd}$) or, equivalently, the disk luminosity ($L_{\rm disk}$) relative to the Eddington luminosity ($L_{\rm Edd}$). Observationally, $L_{\rm disk}/L_{\rm Edd} \approx 1\%$ is usually assumed to be a dividing line between the two accretion modes (e.g., \citealt{Ghisellini2011, HB2014}).

Most ($\approx98\%$) nearby AGNs spend their lives in a low accretion state, making them low-luminosity AGNs (LLAGNs, \citealt{Ho2008, Netzer2013}) which are thought to be powered by hot accretion flows. One of the representative models of hot accretion flows is the advection-dominated accretion flows (ADAFs, \citealt{Ichimaru1977, NY1994, NY1995a, NY1995b}), which is characterized by self-similar solutions with a density profile of $\rho \propto r^{-1.5}$ and a constant mass accretion rate as a function of spherical radius ($r$). Two important properties found in ADAFs are that (i) the flows are convectively unstable and (ii) the Bernoulli parameter of the flow is positive, indicating that strong outflows are a natural outcome of hot accretion flows (e.g., \citealt{NY1994, NY1995a}). These properties led to two variants of ADAF, convection-dominated accretion flow (CDAF, e.g., \citealt{Narayan2000, QG2000, IN2002}) and adiabatic inflow-outflow solution (ADIOS, e.g., \citealt{BB1999, BB2004, Begelman2012}), respectively.

A number of numerical simulations have been performed to better understand the dynamics of hot accretion flows (e.g., \citealt{Stone1999, IA2000, Machida2001, Igumenshchev2003, Pen2003}, see \citealt{Yuan2012b} for a review). One of the most important findings consistently seen in those simulations is that the mass accretion rate decreases with decreasing radius, namely $\dot{M}_{\rm in}(r) \propto r^s$ with $s>0$, or, equivalently, the density profile flatter than the one of ADAF self-similar solutions, i.e., $\rho \propto r^{-q}$ with $q<1.5$. The CDAF model explains the inward decrease of $\dot{M}_{\rm in}$ with large fluxes of both inflowing and outflowing gas in turbulent convective eddies and predicts $s=1$ and $q=0.5$ (e.g., \citealt{Narayan2000, QG2000, IN2002}). In the ADIOS model, the inward decrease of $\dot{M}_{\rm in}$ is due to a genuine mass loss via gas outflows; the model predicts $0<s<1$ and $0.5<q=1.5-s<1.5$ \citep{BB1999, BB2004}. Values of $s=0.4-0.8$ and $q=0.5-1$ were preferentially found in simulations (see \citealt{Yuan2012b} and references therein), which is in general consistent with the ADIOS model. Indeed, both three-dimensional (3D) general relativistic magneto-hydrodynamic (GRMHD) simulations of hot accretion flows \citep{Narayan2012} and 2D simulations of hot accretion flows including magnetic fields \citep{Yuan2012a} showed that hot accretion flows are convectively stable, supporting that hot accretion flows can lose substantial mass via gas outflows (but see \citealt{Bu2016a, Bu2016b} for gas outflows on large spatial scales when the gravitational potential of the nuclear star cluster is included).

Nevertheless, knowledge of the properties of outflows from hot accretion flows has been limited due to the difficulty in tracing the actual outflows by discriminating them from turbulent motions. \cite{Yuan2015} used a ``virtual particle trajectory'' approach and overcame the difficulty in their 3D GRMHD simulations. They found that the outflows from hot accretion flows are dominant in the polar region, while inflows are filling in the equatorial regions, and the geometry of the outflows can be described as conical. Similar results were obtained in another GRMHD simulation in which the collimated and relativistic jet launched from a spinning black hole is surrounded by non-relativistic gas outflows \citep{Sadowski2013}. We clarify the terminology of gas outflows with different physical properties: hereafter, \emph{jet} refers to a highly magnetized, collimated and relativistic gas outflow possibly launched from a spinning black hole \citep{BZ1977} or from the innermost region of an accretion disk \citep{BP1982}, whereas \emph{wind} refers to a moderately magnetized, un-collimated and non-relativistic gas outflow launched from the accretion flow.

Winds have been frequently observed in luminous AGNs for which cold accretion is thought to be operating (e.g., \citealt{Crenshaw2003, Tombesi2010}). However, it is challenging to confirm the presence of winds from hot accretion flows, i.e., in LLAGNs, because the winds are believed to be very hot and generally fully ionized \citep{Yuan2018}. Even though UV and X-ray absorption lines with high outflow velocities have been found in some LLAGNs (e.g., \citealt{Tombesi2014}), due to limited angular resolution it is unclear whether those outflows originate from the accretion flows or from outside regions (e.g., \citealt{CK2012}). Accordingly, there have been attempts to directly determine the radial density profiles of hot accretion flows in a few nearby LLAGNs with X-ray observations. For example, \cite{Wong2011, Wong2014} presented a density profile of NGC 3115 broadely consistent with $\rho \propto r^{-1}$ inside the Bondi radius, within which the gravitational potential energy of the central black hole is larger than the thermal energy of the gas, using Chandra X-ray observations. \cite{Russell2015} showed a similar density profile of $\rho \propto r^{-1}$ for M87 inside the Bondi radius and \cite{Russell2018} found a possible difference between the density profiles in the polar region, i.e., along the jet axis, with $\rho \propto r^{-0.93}$, and in the equatorial region, with $\rho \propto r^{-1.5}$, from Chandra observations. Although these results are consistent with the ADIOS model and possibly indicate the presence of winds in those LLAGNs, they were obtained near the Bondi radius; measurements of density profiles well inside the Bondi radius are needed for a firm conclusion. We note that there are some studies which favor the presence of winds in the supermassive black hole in our Galactic Center, Sagittarius A* (Sgr A*), using spectral energy distribution modelling \citep{Yuan2003}, modelling of the X-ray emission lines \citep{Wang2013}, numerical simulations reproducing the \emph{Fermi} Bubbles possibly inflated by those winds \citep{Mou2014}, and modelling of the motion of the gas cloud G2 slowed down by a drag force \citep{Gillessen2018}.

Winds have important astrophysical implications. The actual rate of mass accreted onto the black hole could be substantially smaller than the accretion rate measured through X-ray observations at the Bondi radius (Bondi accretion rate, $\dot{M}_{\rm Bondi}$) due to the mass loss via winds. Therefore, a major factor in the faintness of LLAGNs might be the reduced mass accretion rate \citep{Bower2003}, not a very low radiative efficiency as usually assumed \citep{XY2012}. Also, rotational energy of spinning black holes must be extracted efficiently to explain the observed high kinetic jet powers with small mass accretion rate \citep{NT2015}. Furthermore, winds have a large cross section and may regulate star formation in the host galaxies via momentum transfer \citep{Yoon2018, Yuan2018}.

\begin{deluxetable*}{ccccc}[t!]

\tablecaption{Summary of VLBA archive data used in this study}
\tablehead{
\colhead{Project code} & \colhead{Obs. date} &
\colhead{Frequency [GHz]} &
\colhead{D Term cal.} & \colhead{EVPA cal.} \\
(1) & (2) & (3) & (4) & (5)
}
\startdata
 BJ020A & 1995 Nov 22 & 8.11, 8.20, 8.42, 8.59 & OQ 208 & OJ 287 (UMRAO)\\
 BJ020B & 1995 Dec 09 & 4.71, 4.76, 4.89, 4.99 & OQ 208 & 3C 273 (UMRAO)\\
 BC210B & 2013 Mar 09 & 4.85, 4.88, 4.92, 4.95, 4.98, 5.01, 5.04, 5.08 & M87 & N/A\\
 BC210C & 2014 Jan 29 & 4.85, 4.88, 4.92, 4.95, 4.98, 5.01, 5.04, 5.08 & M87 & N/A\\
 BC210D & 2014 Jul 14 & 4.85, 4.88, 4.92, 4.95, 4.98, 5.01, 5.04, 5.08 & M87 & N/A\\
 BH135F & 2006 Jun 30 & 1.65, 1.66, 1.67, 1.68 & M87 & 3C 286\\
 BC167C & 2007 May 28 & 1.65, 1.66, 1.67, 1.68 & M87 & 3C 286\\
 BC167E & 2007 Aug 20 & 1.65, 1.66, 1.67, 1.68 & M87 & 3C 286
\enddata
\tablecomments{(1) Project code of VLBA observations. (2) Observation date. (3) Observing frequency for all sub-bands. (4) Source used for calibration of instrumental polarization. (5) Source used for EVPA calibration. `(UMRAO)' means that we corrected the EVPA by comparing the VLBI integrated EVPAs with the EVPAs obtained from contemporaneous single dish observations by the University of Michigan Radio Astronomy Observatory. N/A implies that EVPA calibration was not available. 3C 286 has a stable integrated EVPA of $33^{\circ}$ at the frequencies of our interest \citep{PB2013}.\label{Info}}

\end{deluxetable*}

Another important role played by winds is their effect on the collimation of AGN jets. It has been a long-standing problem how jets in AGNs can be highly collimated and accelerated to nearly the speed of light. It is widely accepted that the acceleration and collimation zone in AGN jets are co-spatial and located within about $10^5$ Schwarzschild radii ($r_{\rm s}$, \citealt{Marscher2008}). MHD models predict that magnetic fields can accelerate AGN jets to relativistic speeds if the jets are systematically collimated (e.g., \citealt{Vlahakis2015}). It is difficult for the jets to be confined by themselves (e.g., \citealt{Eichler1993, BL1994, Komissarov2007}) and an external confining medium is necessary to produce the observed highly collimated jets. Previous theoretical studies suggest that winds are the primary candidates for this medium \citep{TB2002, MG2004, BT2005, DeVilliers2005, Gracia2005, GL2016, Nakamura2018}. 


M87 serves as a unique laboratory for studying AGN jets and their formation, collimation, and acceleration thanks to its proximity with a distance of 16.7 Mpc \citep{Mei2007} and its extremely massive black hole with a mass of $M_{\rm BH} = (3.5-6.6)\times10^9M_{\odot}$ \citep{GT2009, Gebhardt2011, Walsh2013}. Accordingly, this source has been studied extensively especially on scales corresponding to the jet acceleration and collimation zone. One of the most notable results is the discovery of an edge-brightened jet structure with a systematic collimation of the jet on scales $\gtrsim100\ r_{\rm s}$ \citep{Junor1999}. The large-scale collimation profile shows a transition from a semi-parabolic jet with $z \propto R^{1.7}$, where $z$ and $R$ denote the jet distance and the jet radius, respectively, to a conical jet at a transition location near the Bondi radius \citep{AN2012}. The precise constraint on the location of the black hole by core-shift analysis \citep{Hada2011} together with the source size measured with the Event Horizon Telescope (EHT) at 1.3 mm \citep{Doeleman2012} allowed to constrain the innermost collimation profile. The profile is consistent with a parabolic geometry \citep{NA2013} but shows indication of a slight deviation from the larger scale profile (\citealt{Hada2013}, see also \citealt{Hada2016, Mertens2016, Kim2018a, Walker2018}). 

There has been growing evidence for gradual acceleration of the jet inside the Bondi radius as well, though the scale on which bulk jet acceleration occurs is a matter of debate. Observations of HST-1, a peculiar feature that consists of a quasi-stationary component from which superluminal components are emerging and is the location of the multiwavelength flare observed around 2005 \citep{Cheung2007}, show superluminal motions with velocities larger than $6c$ (with $c$ being the speed of light) at optical wavelengths \citep{Biretta1999}, and with velocities of $\approx4c$ at radio wavelengths \citep{Cheung2007, Giroletti2012}. \cite{Asada2014} found a systematic acceleration of the jet at a distance of $\approx10^5\ r_{\rm s}$, supported by the slow velocities obtained on smaller scales with the Very Long Baseline Array (VLBA) at 15 GHz \citep{Kovalev2007}. However, as already noted in \cite{Kovalev2007}, the observed one-sideness of the jet at a distance of only $\approx3$ milliarcseconds from the radio core is difficult to explain with sub-luminal motions at the same distance. Other studies suggest that the jet acceleration occurs on a much smaller scale \citep{Mertens2016, Hada2017, Walker2018} and constraining the acceleration profile at various jet distances is still on-going (Park et al. 2018, in prep.).

The observation of jet collimation and acceleration on the same spatial scales is consistent with the scenario that the jet is collimated by an external medium with a relatively shallow pressure profile, which results in gradual acceleration of the jets in an MHD process \citep{Komissarov2009, Lyubarsky2009}. However, it has not been possible to either probe the external medium with observations or to verify the general picture of jet collimation and acceleration. In this study, we investigate Faraday rotation, the rotation of the plane of linear polarization by intervening magnetic fields, in the jet of M87. When linearly polarized emission passes through a magnetized medium, Faraday rotation occurs. The amount of rotation of the electric vector position angle (EVPA), $\Delta\chi$, is related to the Faraday rotation measure (RM) via $\Delta\chi = \rm RM\lambda^2$, where $\lambda$ is the wavelength. RM is proportional to the integral of the product of free electron density ($n_e$) and line of sight component of the magnetic field ($B$) along the path from emitter to observer ($l$), meaning ${\rm RM} \propto \int n_e(l)B(l)dl$ (e.g., \citealt{GW1966}). Thus, observations of the Faraday rotation of polarized jets can probe the magnetized medium between the jet and the observer, i.e., the external medium. Unfortunately, the jets in nearby LLAGNs are usually very weakly polarized (see e.g., \citealt{Bower2017} for more discussion) and the Faraday rotation observations have been limited to specific emitting regions in some sources such as Sgr A* (e.g., \citealt{Bower2003, Bower2018, Marrone2006, Marrone2007, Liu2016}), 3C 84 (e.g., \citealt{Taylor2006, Plambeck2014, Nagai2017, Kim2018b}), and M87 (e.g., \citealt{ZT2002, Kuo2014}). In this work, we obtain RM values at various locations in the M87 jet by exploiting multifrequency VLBA data from multiple epochs and present the radial RM profile of the jet between 5,000 and 200,000 $r_{\rm s}$. Then, we test the conjecture that winds are launched from hot accretion flows and serve as the external confining medium of the jet using the RM data.


\begin{figure*}[!t]
\begin{center}
\includegraphics[trim=0mm 0mm 0mm 0mm, clip, width = 179mm]{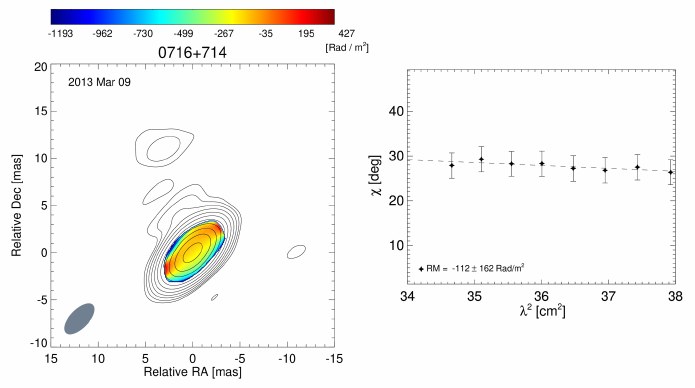}
\caption{\emph{Left:} Color map of the distribution of RM overlaid on contours of the total intensity of the calibrator 0716+714 in the BC210B session observed at 5 GHz. The colorscale of RM in units of $\rm rad/m^2$ is shown at the top. The beam size is illustrated by the gray shaded ellipse. Contours start at 0.79 mJy per beam and increase by factors of 2. \emph{Right:} EVPA as function of $\lambda^2$ at the center of the map shown in the left panel. The dashed line is the best-fit $\lambda^2$-law with $\rm RM=-112\pm162\ rad/m^2$.\label{calRM}}
\end{center}
\end{figure*}

\section{Archival data and data reduction}
\label{reduction}

We searched the VLBA archive for data suitable for a study of linear polarization and Faraday rotation in the M87 jet. We selected those data in which (i) different sub-bands are sufficiently separated in wavelength, (ii) both parallel and cross-hand visibilities are available, and (iii) M87 is observed as a primary target in full-track observing mode. Using these criteria, we are left with one data set at 8 GHz, four data sets at 5 GHz, and many data sets at 2 GHz. We note that there are multifrequency VLBA data obtained quasi-simultaneously in 7 different sub-bands from 8.1 and 15.2 GHz in the literature \citep{ZT2002} which we could not find in the VLBA archive. Therefore, these data are not included in our analysis but we show that our results are consistent with the results of their work in Section~\ref{sectvar}. We found that the distribution of RM in the jet in different data sets at 2 GHz are more or less the same and chose three data sets among them for which all 10 VLBA antennas are available and the weather was good. We show the list of the eight VLBA archive data sets we analyzed and the basic information for each observation in Table~\ref{Info}. In total, we analyzed eight different polarization data sets of M87 taken by the VLBA (one at 8 GHz, four at 5 GHz, and three at 2 GHz). 

A standard data post-correlation process was performed with the National Radio Astronomy Observatory’s (NRAO) Astronomical Image Processing System (AIPS, \citealt{Greisen2003}). We corrected ionospheric dispersive delays using the ionospheric model provided by the Jet Propulsion Laboratory, antenna parallactic angles, and instrumental delays using scans on bright calibrators. Amplitude calibration was performed by using the antenna gain curves and system temperatures with an opacity correction. We performed global fringe fitting with a solution interval between 10 and 30 seconds assuming a point source model. Bandpass calibration was performed by using scans on bright calibrators. The cross-hand R-L phase and delay offsets were calibrated by using scans on bright calibrators. We used the Caltech Difmap package \citep{Shepherd1997} for imaging and phase and amplitude self-calibration. We determined the feed polarization leakage (D-terms) for each antenna and for each sub-band by using the task LPCAL \citep{Leppanen1995} in AIPS with a total intensity model of the D-term calibrators. We used OQ 208 or M87 for the D-term correction (Table~\ref{Info}) because of their very low degree of linear polarization (usually $\lesssim 1\%$). 

\begin{figure*}[!t]
\begin{center}
\includegraphics[trim=10mm 19mm 7mm 9mm, clip, width = \textwidth]{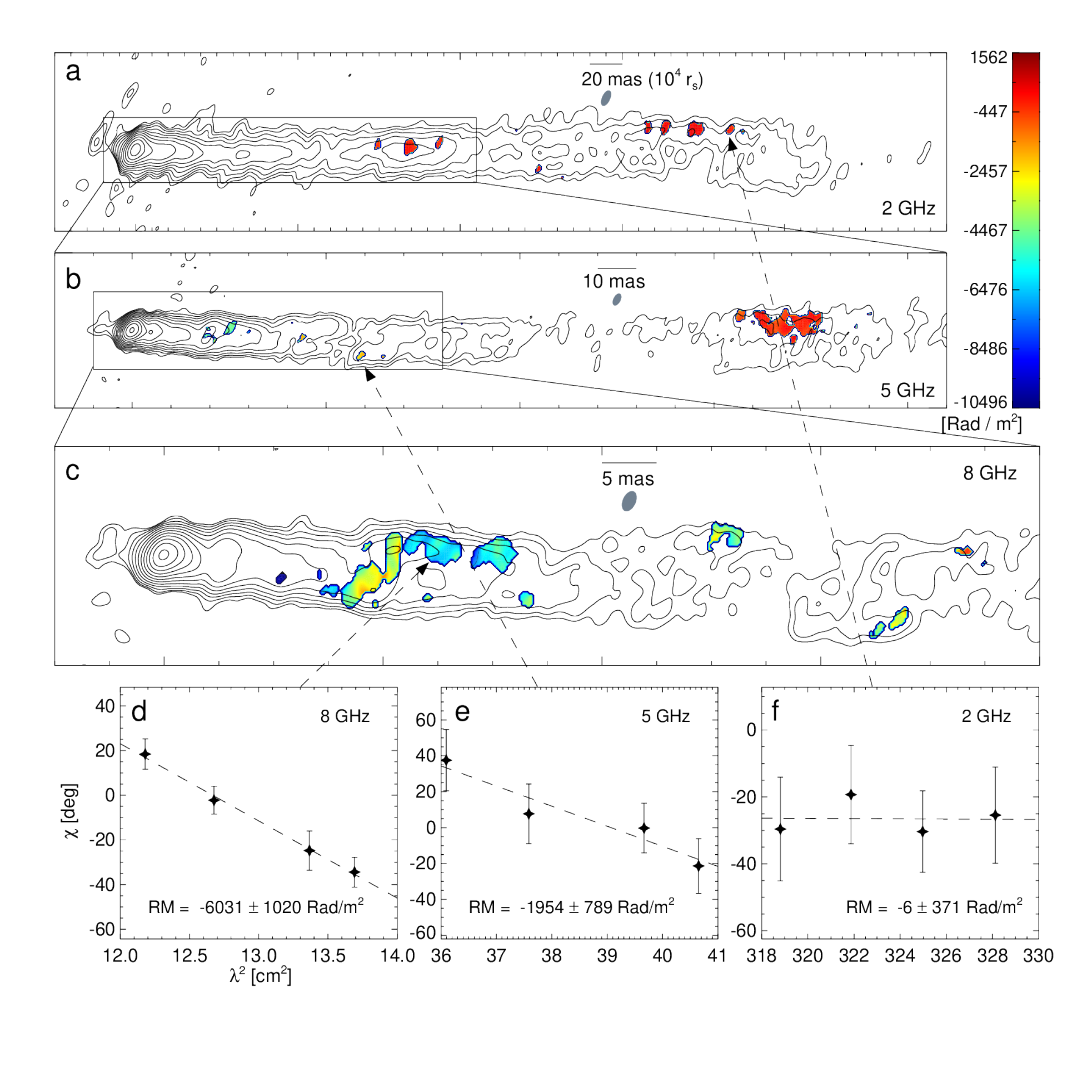}
\caption{Colormap of the RM distribution, overlaid on contours of the total intensity of the M87 jet in three VLBA data sets (out of eight) at 2 (\textbf{a}), 5 (\textbf{b}), and 8 GHz (\textbf{c}). Contours start at 0.79, 0.54, and 0.53 mJy per beam for the 2 GHz, 5 GHz, and 8 GHz maps, respectively, and increase by factors of 2. The RM colorscale in units of $\rm rad/m^2$ is shown at the top-right corner. Beam sizes are illustrated by the gray shaded ellipses. All maps are rotated clockwise by $23^\circ$ relative to astronomical R.A.--Dec coordinates for better visualization. EVPA as function of $\lambda^2$, along with the best-fit $\lambda^2$ laws at the locations indicated by the black dashed arrows, is shown in \textbf{d--f}. We note that all RMs measured at different locations show good $\lambda^2$ fits (see Figure~\ref{extendedRM}). We omitted the jet and RMs at $\approx 900$ mas from the core at 2 GHz for better visualization (see Figure~\ref{extendedRM}).\label{RM}}
\end{center}
\end{figure*}

\begin{figure*}[!t]
\begin{center}
\includegraphics[trim=0mm 34mm 2mm 25mm, clip, width=\textwidth]{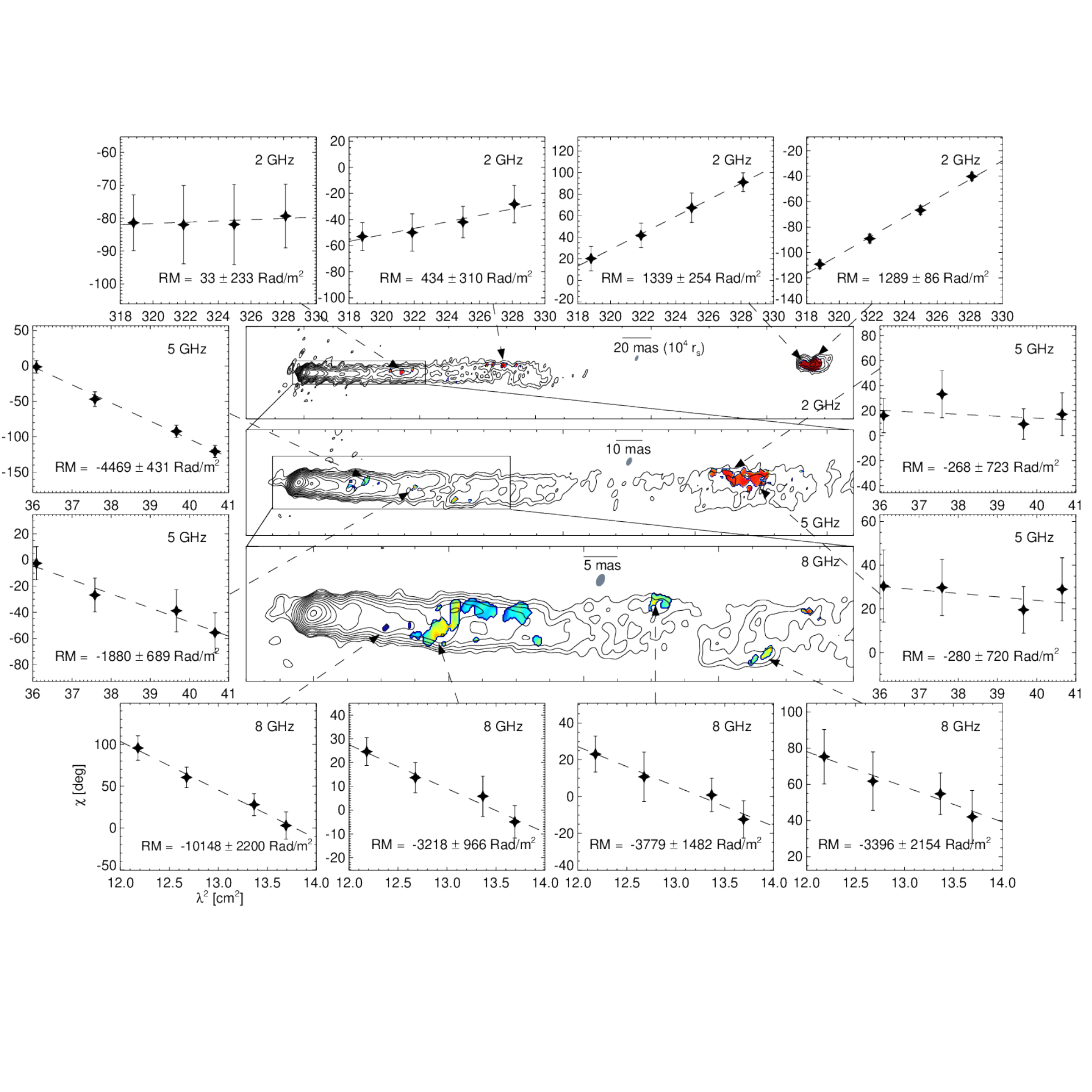}
\caption{Same as Figure~\ref{RM} but including HST-1 at $\approx 900$ mas from the core at 2 GHz and showing EVPA--$\lambda^2$ diagrams for various jet locations where significant Faraday rotation is detected. \label{extendedRM}}
\end{center}
\end{figure*}

The EVPA calibration was performed by comparing the integrated EVPAs of the VLBI maps of the calibrators with the EVPAs obtained in contemporaneous single dish polarization observations of the University of Michigan Radio Astronomy Observatory (UMRAO), or by using 3C 286 for which a stable integrated EVPA of $\approx33^{\circ}$ is known at the frequencies of our interest \citep{PB2013}, if available. However, we note that EVPA calibration is not critical for our purpose because the expected amount of EVPA correction for different sub-bands is almost the same. For example, we present the RM map and EVPA as a function of $\lambda^2$ at the map center of one of the calibrators in BC210B session, 0716+714, in Figure~\ref{calRM}. Even though we could not perform EVPA correction for this epoch (see Table~\ref{Info}), the difference in EVPAs in different sub-bands is much smaller than the error bars and the obtained RM value is consistent with the previous measurements with the VLBA \citep{Hovatta2012}. We check the RM of the calibrators in all the data we analyzed to ensure that the detected RM for M87 is not due to potential errors in EVPA calibration but is intrinsic to the source itself.

We obtained RM values at various positions in the M87 jet from measuring EVPAs in different sub-bands (intermediate frequencies) in each dataset (see Table~\ref{Info}). We considered four error sources in EVPA: random error, systematic error induced by imperfect CLEAN procedures, by imperfect D-term calibration, and by imperfect EVPA calibration. We present the details of error estimation in Appendix~\ref{appendixa}. For obtaining RM maps, we first convolved the maps in different sub-bands with the synthesized beam of the sub-band at the lowest frequency. Then, we fitted a linear function to the EVPAs from different sub-bands versus $\lambda^2$ for each pixel where the linear polarization intensity exceeds $1.5\sigma$ in all sub-bands, with $\sigma$ being the full uncertainty including D-term errors and CLEAN errors \citep{Hovatta2012}. We discuss the significance levels of the observed RMs in Appendix~\ref{appendixb}. We fitted the EVPA data several times including potential $n\pi$ rotations and used the fit that provided us with the lowest $\chi^2$ value.

\begin{figure*}[!t]
\begin{center}
\includegraphics[trim=22mm 18mm 10mm 3mm, clip, width = 135mm]{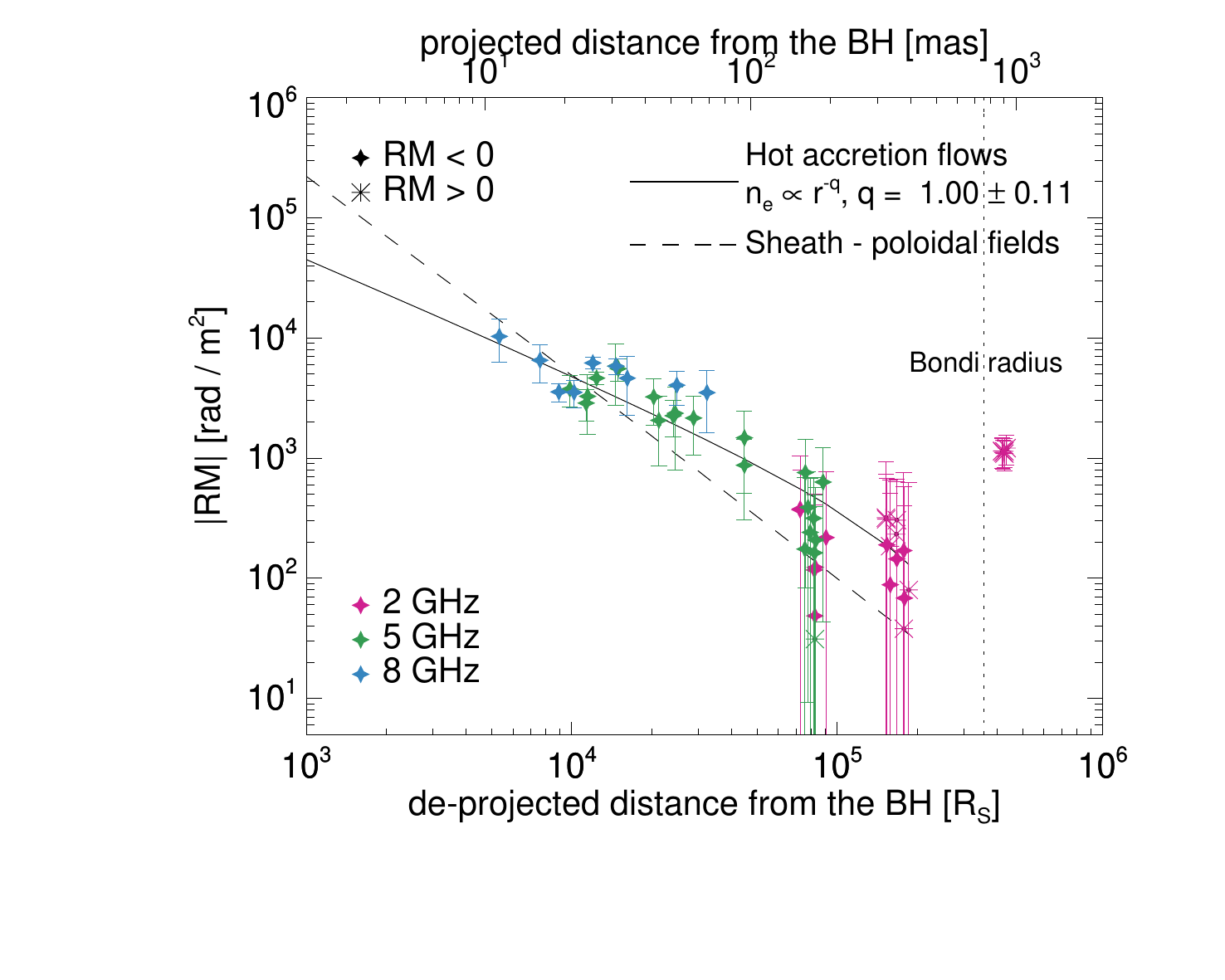}
\caption{Absolute values of RM as function of de-projected distance from the black hole (bottom abscisa) and projected angular distance (top abscissa). The red, green, and blue data points are obtained at 2, 5, and 8 GHz, respectively. Diamonds and asterisks denote negative and positive RMs, respectively. The vertical dotted line indicates the Bondi radius \citep{Russell2015}. The solid (dashed) line is the best-fit function of the hot accretion flows (the sheath) model to the data points (see Section~\ref{sectscreen}). The hot accretion flows model describes the observed data better than the sheath model (see Section~\ref{sectcomp}). The electron density $n_e$ scales like $n_e \propto r^{-q}$ with $q=1.00\pm0.11$ in the best-fit function of the hot accretion flows model. All RM values displayed here were obtained after subtracting 130 $\rm rad/m^2$ from our measurement results; the RM errors were obtained after adding 300 $\rm rad/m^2$ in quadrature to our measurement uncertainties (see Section~\ref{sectoutside}). \label{radial}}
\end{center}
\end{figure*}

\section{Analysis and Results}

\subsection{RM maps}
\label{sectmaps}

In Figure~\ref{RM}, we present example RM maps overlaid on the total intensity distribution of the jet for one observation at each frequency. The EVPA as function of $\lambda^2$ at three different locations of the jet is shown with good $\lambda^2$ fits. We obtained good $\lambda^2$ fits for the other RMs measured at different locations as well (some of them are shown in Figure~\ref{extendedRM}) and also in the other five data sets not presented in Figure~\ref{RM}. We omitted the jet and RMs at $\approx 900$ mas from the core at 2 GHz for better visualization; those data are presented in Figure~\ref{extendedRM}. At lower frequencies, Faraday rotation is observable in more outward regions of the jet due to longer cooling times of the jet plasma. At higher frequencies, Faraday rotation is observable closer to the compact upstream emission thanks to better angular resolution and less depolarization. We note that the RM distributions are patchy at all frequencies because significant linear polarization is detected only in some parts of the jet, possibly due to substantial de-polarization in the other parts. We also note that it is unlikely that those patchy RMs are artifacts because we found that the RMs in different epochs at the same observing frequency are detected in similar locations of the jet (see Appendix~\ref{sectstack}).

\subsection{Radial RM profile}
\label{sectradial}

To obtain a radial RM profile along the jet, we calculated spatially binned RM by taking the weighted mean of all values in each separated region of the map with similar RM values. A priori, taking a weighted mean over a part of a map assumes that all individual pixels are independent from each other, which is not the case here. Pixels values are correlated across the extension of a resolution element (here, the synthesized beam). Thus, we first calculated a mean value, then its formal error (which assumes all pixels to be uncorrelated), and then multiplied this formal error by $\sqrt{n\Sigma_{\rm FWHM}/\Sigma_{\rm RM}}$, where $n$ is the number of the pixels used for taking the mean, $\Sigma_{\rm RM}$ the size of the map region with RM values, and $\Sigma_{\rm FWHM}$ the area within the full width at half maximum of the synthesized beam. We present the mean distance from the black hole, the binned RM values, and corresponding RM errors in Table~\ref{RMvalue}. This data is used for our further analysis.

\begin{figure*}[!t]
\begin{center}
\includegraphics[trim=22mm 18mm 12mm 3mm, clip, width = 88mm]{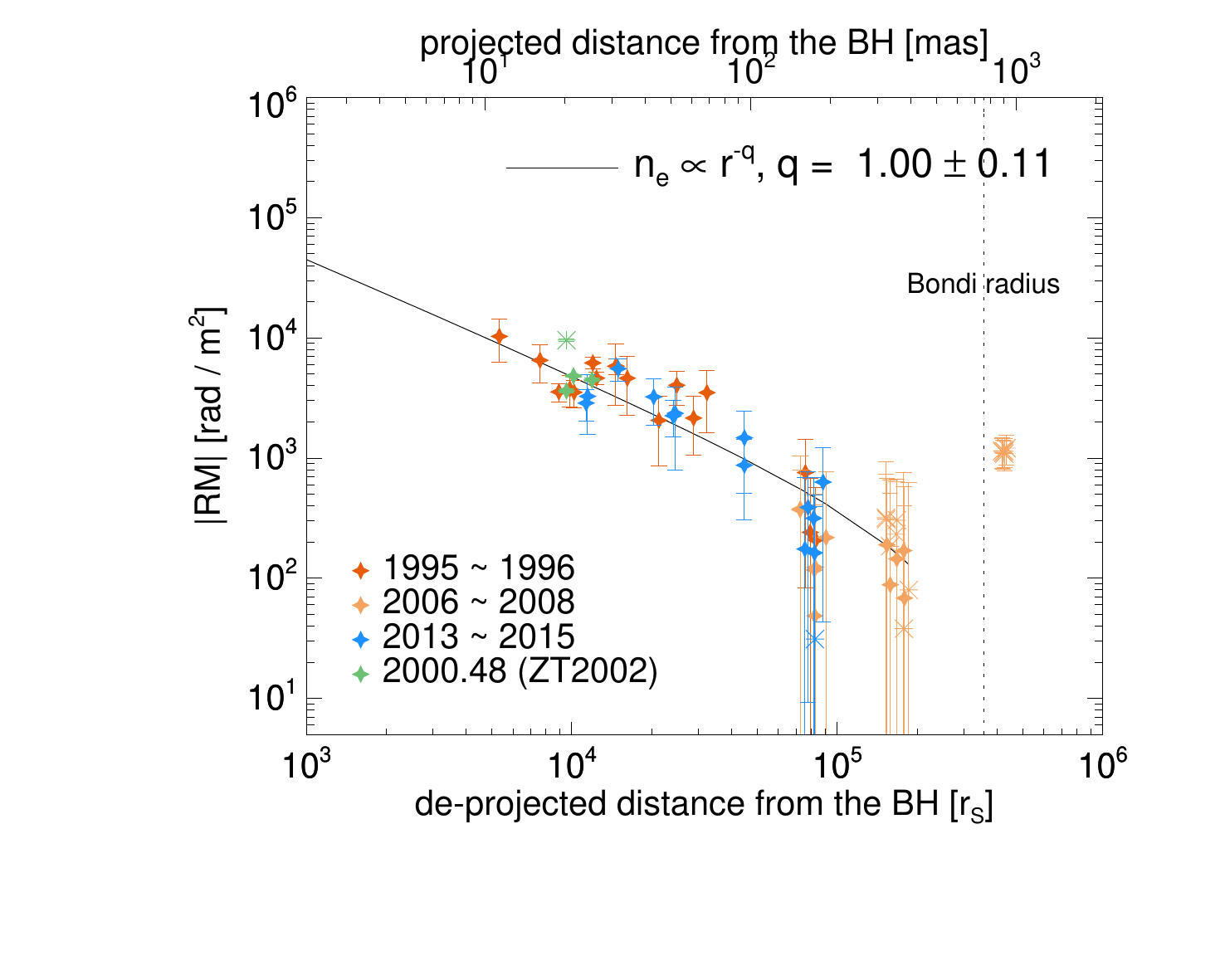}
\includegraphics[trim=22mm 18mm 12mm 3mm, clip, width = 88mm]{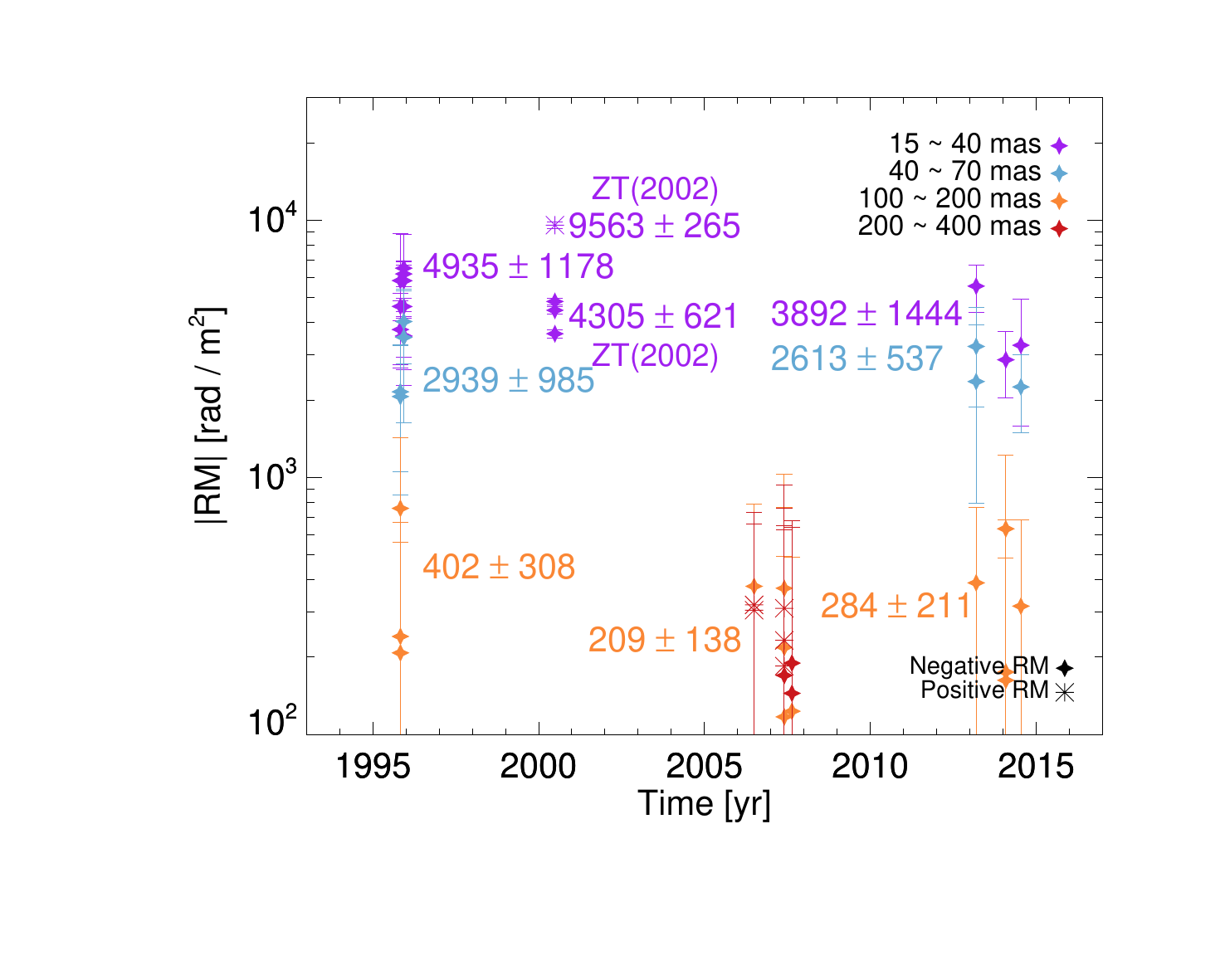}
\caption{\emph{Left:} Same as Figure~\ref{radial} but with data points obtained in different epochs shown in different colors. \emph{Right:} Absolute value of RM as a function of time. Values obtained at different angular distance ranges from the black hole are shown in different colors. The mean and standard deviation of RM values for each group (obtained in each time period for each jet distance) are noted next to the data points. The diamonds and asterisks denote negative and positive RMs, respectively. The data points obtained in the middle of 2000 are from a previous RM study of the M87 jet \citep{ZT2002}. \label{temporal}}
\end{center}
\end{figure*}

In Figure~\ref{radial}, we present the absolute values of RM as function of de-projected distance from the black hole in units of $r_{\rm s}$. We assumed a black hole mass of $M_{\rm BH} = 6\times10^9M_\odot$ \citep{Gebhardt2011}, a viewing angle of 17 deg\footnote{We note that the viewing angle of the M87 jet is a matter of on-going discussion. Some studies suggest relatively large angles of $\theta\gtrsim30^\circ$ (e.g., \citealt{Owen1989, Ly2007, Hada2016}), while other studies reported rather small viewing angles of $\theta\lesssim19^\circ$ (e.g., \citealt{Biretta1999, WZ2009, Perlman2011, Mertens2016, Walker2018}). In this study, we adopt a viewing angle $17^\circ$ based on the results of \cite{Mertens2016} and consideration of the upper limit of $\theta\lesssim19^\circ$ derived from the velocity measurement at HST-1 \citep{Biretta1999}, as in \cite{Walker2018}.} \citep{Mertens2016}, and a distance between black hole and radio core as estimated by core-shift analysis of the M87 jet \citep{Hada2011} to convert the observed projected jet distance from the radio cores into the de-projected distance from the black hole. Remarkably, the RM decreases systematically along the jet over nearly two orders of magnitude in distance (from 5,000 to 200,000 $r_{\rm s}$) inside the Bondi radius ($3.6\times10^5r_{\rm s}$; \citealt{Russell2015}). Our results substantially improved the radial RM profile of the M87 jet that was previously limited to a specific jet location at $\approx20$ mas from the core obtained in the pioneering RM study of the M87 jet \citep{ZT2002}. The sign of the rotation measure is preferentially negative inside the Bondi radius except in the outer jet region (at distance of $\approx2\times10^5\ r_{\rm s}$) where RM errors are comparable to the RM values, which makes the RM sign ambiguous. However, at the location of HST-1 (at $\approx4\times10^5\ r_{\rm s}$), the observed RMs suddenly increase by a factor of $\approx10$ compared with those at $\approx2\times10^5\ r_{\rm s}$ and their signs are always positive, which is opposite to the signature observed in the inner jet region (Figure~\ref{radial}). This result is in good agreement with the previous measurements by the Very Large Array (VLA) observations \citep{Chen2011}. Since we focus on the behavior of RMs inside the Bondi radius in this paper, we briefly discuss the results of RMs at HST-1 in Section~\ref{sectHST} and more detailed results will be presented in a forthcoming paper (Park et al. 2018, in prep.).

\subsection{Contribution of RM sources outside the Bondi radius}
\label{sectoutside}

We investigate the source of RMs inside the Bondi radius in this paper. However, there are three candidates other than gas within the Bondi radius which can contribute to the observed RM values: the Galactic interstellar medium (ISM), the intergalactic medium (IGM) in the Virgo cluster, and the diffuse gas not bound by the black hole's gravity in M87. The Galactic ISM would contribute less than $\approx20\rm\ rad/m^2$ because of the large galactic latitude of $b=74.5^{\circ}$ for M87 \citep{Taylor2009}. The IGM in the Virgo cluster is expected to contribute less than $\approx30\rm\ rad/m^2$, based on the RM observations of other galaxies in the cluster \citep{Wezgowiec2012}. However, the contribution of the diffuse gas in M87 outside the Bondi radius would not be negligible. Previous VLA observations of M87 showed that RMs of the larger scale jet outside the Bondi radius are typically $\approx130\rm\ rad/m^2$ but values as low as $\approx-250\rm\ rad/m^2$ and as high as $\approx650\rm\ rad/m^2$ are also seen in some parts of the jet \citep{Owen1990, Algaba2016}. Therefore, we subtracted $130\rm\ rad/m^2$ from our observed RM values and added $300\rm\ rad/m^2$ to the RM errors quadratically, which is used in Figure~\ref{radial} and for our further analysis.

\subsection{Variability}
\label{sectvar}

Our data are obtained in different periods from 1996 to 2014, so RM variability might affect the results. We also included the results of a previous study of RM of the M87 jet \citep{ZT2002} for investigating potential RM variability. One can divide our data into four time groups, obtained in 1995--1996, 2000.48, 2006--2008, and 2013--2015. We show the absolute values of RM from different groups with different colors as a function of distance from the black hole in the left panel of Figure~\ref{temporal}. The data obtained in different periods do not show significant deviation from each other. We also present the RM values as a function of time obtained in four different jet distance ranges, 15--40, 40--70, 100--200, and 200--400 mas, with different colors (the right panel of Figure~\ref{temporal}). The mean values from different groups in the same jet distance range are consistent with each other within $1\sigma$ in almost all cases, suggesting that there is no significant temporal variability in RM. However, one exception is the case of positive RMs detected at $\approx25$ mas from the radio core in 2000.48 presented in the literature \citep{ZT2002}. This value is larger than others obtained at similar jet distance by a factor of $\approx2$ and its sign is opposite. It is reasonable to consider that the positive RMs might be locally transient and not related to a global behavior of RM of the M87 jet because (i) the region of positive RMs is much smaller than that of negative RMs at a similar jet distance by a factor of $\approx20$ \citep{ZT2002} and (ii) positive RMs are not detected in other epochs and at other jet distances except in the outer jet region where RM errors are comparable to the RM values, which makes the RM sign ambiguous.

\begin{deluxetable}{cccc}[t!]

\tablecaption{Binned rotation measure values}
\tablehead{
\colhead{Session} & \colhead{Proj. dist. [mas]} &
\colhead{RM $\rm [rad/m^2]$} & \colhead{$\sigma_{\rm RM}\rm\ [rad/m^2]$}\\
(1) & (2) & (3) & (4)
}
\startdata
\centering \multirow{10}{0.2\columnwidth}{\centering BJ020A} & 10.87 & -10163.47 & 3965.69 \\
& 15.55 & -6381.18 & 2278.09 \\
& 18.35 & -3421.25 & 541.23 \\
& 20.97 & -3391.10 & 849.69 \\
& 24.69 & -6054.02 & 615.75 \\
& 30.23 & -5688.09 & 804.28 \\
& 33.37 & -4489.37 & 2327.39 \\
& 51.38 & -3908.49 & 1239.84 \\
& 66.79 & -3374.41 & 1852.78 \\
\hline
\centering \multirow{8}{0.2\columnwidth}{\centering BJ020B} & 20.07 & -3628.40 & 1049.11 \\
& 25.39 & -4495.79 & 472.80 \\
& 30.05 & -5698.63 & 3064.56 \\
& 43.83 & -1932.51 & 1169.71 \\
& 59.31 & -2022.79 & 1056.72 \\
& 156.90 & -627.96 & 603.85 \\
& 163.58 & -110.87 & 312.19 \\
& 171.72 & -77.88 & 194.93 \\
\hline
\centering \multirow{6}{0.2\columnwidth}{\centering BC210B} & 30.73 & -5414.65 & 1117.57 \\
& 41.89 & -3101.13 & 1317.59 \\
& 50.59 & -2229.21 & 1535.19 \\
& 92.36 & -742.15 & 478.03 \\
& 160.65 & -259.05 & 232.85 \\
& 170.49 & 161.17 & 201.12 \\
\hline
\centering \multirow{5}{0.2\columnwidth}{\centering BC210C} & 23.18 & -2736.92 & 779.55 \\
& 92.35 & -1345.83 & 919.99 \\
& 156.32 & -45.10 & 414.45 \\
& 170.03 & -32.60 & 136.35 \\
& 183.02 & -501.42 & 505.80 \\
\hline
\centering \multirow{3}{0.2\columnwidth}{\centering BC210D} & 23.47 & -3135.18 & 1653.81 \\
& 49.63 & -2119.48 & 693.08 \\
& 168.70 & -185.79 & 221.31 \\
\hline
\centering \multirow{7}{0.2\columnwidth}{\centering BH135F} & 149.44 & -247.30 & 288.74 \\
& 170.00 & 81.37 & 198.19 \\
& 315.35 & 449.35 & 285.42 \\
& 347.12 & 434.49 & 196.92 \\
& 370.33 & 61.64 & 137.53 \\
& 873.80 & 1242.69 & 36.47 \\
& 899.69 & 1340.05 & 143.18 \\
\hline
\centering \multirow{9}{0.2\columnwidth}{\centering BC167C} & 149.35 & -240.64 & 592.94 \\
& 169.17 & 12.77 & 233.97 \\
& 187.30 & -88.17 & 460.41 \\
& 316.27 & 439.96 & 547.68 \\
& 326.85 & 314.65 & 358.40 \\
& 345.39 & 362.52 & 262.42 \\
& 368.23 & -39.88 & 508.98 \\
& 865.47 & 1270.15 & 102.53 \\
& 883.03 & 1269.38 & 44.13 \\
\hline
\centering \multirow{8}{0.2\columnwidth}{\centering BC167E} & 170.44 & 6.70 & 217.93 \\
& 317.98 & -59.62 & 388.00 \\
& 327.55 & 41.67 & 292.40 \\
& 346.71 & -14.73 & 394.56 \\
& 368.51 & 167.98 & 446.63 \\
& 384.56 & 209.87 & 460.24 \\
& 866.06 & 1276.49 & 117.15 \\
& 883.04 & 1221.10 & 48.23
\enddata
\tablecomments{(1) Project code of VLBA observations. (2) Mean projected distance from the black hole of the region where the RMs are measured, in units of milliarcseconds. (3) Binned RM values in units of $\rm rad/m^2$. (4) $1\sigma$ errors of the binned RMs. All RM values are those before subtracting 130 $\rm rad/m^2$ and the RM errors are before adding 300 $\rm rad/m^2$ in quadrature to the uncertainties (see Section~\ref{sectoutside}).\label{RMvalue}}
\end{deluxetable}

\subsection{The Faraday screen}
\label{sectscreen}

\subsubsection{Internal Faraday rotation and depolarization}
\label{sectinternal}

If the Faraday rotating electrons are intermixed with the synchrotron emitting jet plasma, internal Faraday rotation can occur. \cite{Burn1966} showed that the complex polarization ($\mathcal{P}$) of a synchrotron-emitting uniform slab with a purely regular magnetic field (see \citealt{Sokoloff1998} for the case of a non-uniform or an asymmetric slab) is given by
\begin{equation}
\mathcal{P} \equiv Q + iU = p_0I\frac{\sin \phi\lambda^2}{\phi\lambda^2}{\rm e}^{2i(\chi_0 + \frac{1}{2}\phi\lambda^2)},
\label{burneq}
\end{equation}
where $Q$, $U$, and $I$ are intensity in Stokes $Q$, $U$, and $I$ maps, respectively, $p_0$ is the intrinsic fractional polarization, $\chi_0$ the intrinsic EVPA, and $\phi$ the Faraday depth. However, internal Faraday rotation in sources with more realistic geometries and magnetic field structures usually results in deviation from a $\lambda^2$ law after total rotations $\gtrsim45^\circ$ \citep{Burn1966, Sokoloff1998, Homan2012}.

\begin{figure}[!t]
\begin{center}
\includegraphics[trim=3mm 5mm 10mm 2mm, clip, width = 86mm]{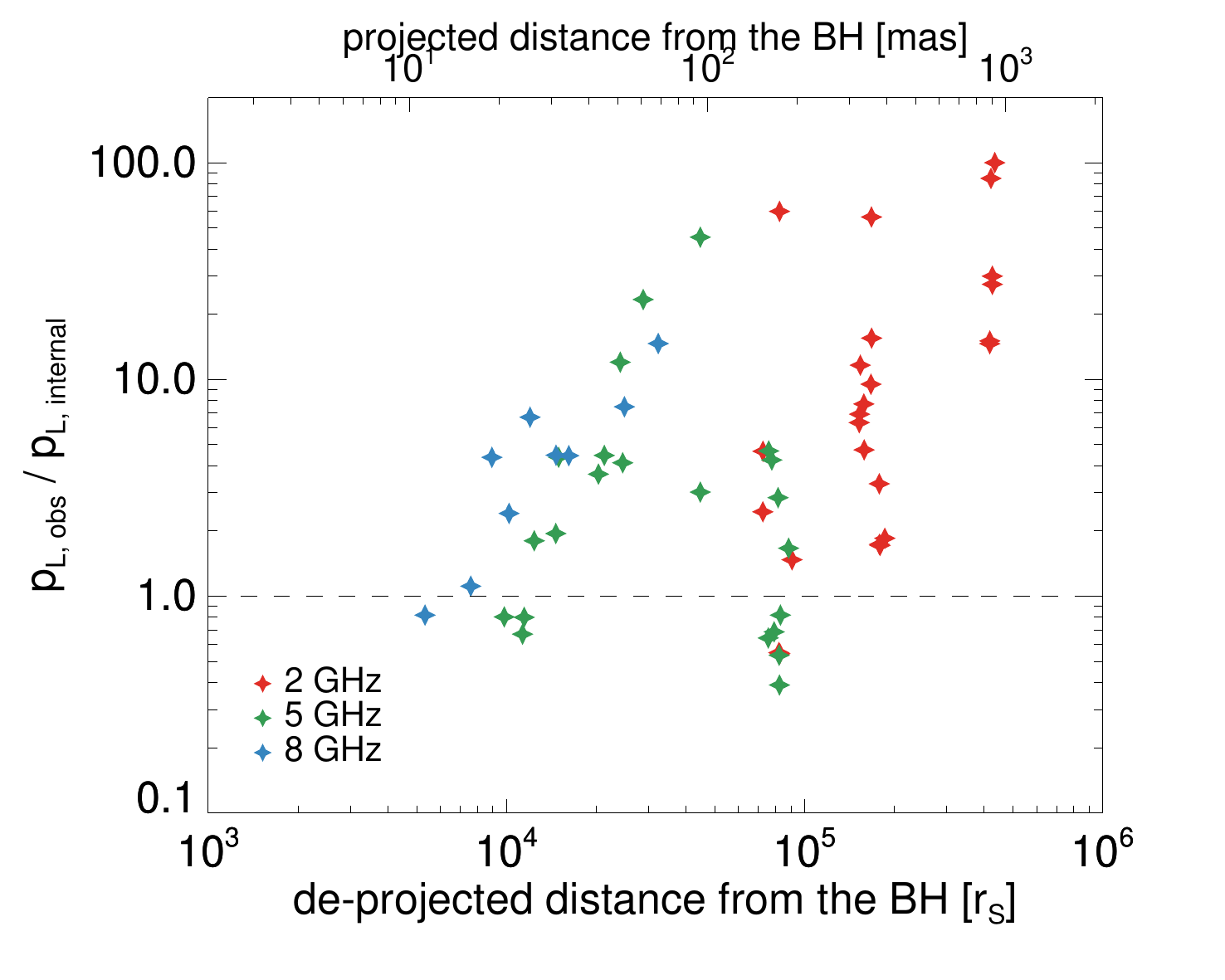}
\caption{Ratio of the observed degree of linear polarization to the expected one in the internal Faraday rotation model (Equation~\ref{burneq}, \citealt{Burn1966}) as a function of distance from the black hole. The horizontal dashed line shows unity ratio. \label{internalfrac}}
\end{center}
\end{figure}

We tested whether the observed degree of polarization and Faraday rotation can be explained with Equation~\ref{burneq} or not. We compared the degree of linear polarization expected in this model, $p_{\rm L, internal} = p_0|\rm sinc(2{\rm RM}\lambda^2)|$, with the observed one, $p_{\rm L, obs}$. We assumed $p_0 \approx 0.75$ because this is the maximum allowed degree of linear polarization for optically thin synchrotron radiation \citep{Pacholcyzk1970}. In Figure~\ref{internalfrac}, we present $p_{\rm L, obs} / p_{\rm L, internal}$ as a function of de-projected distance from the black hole. Most of the data points are much larger than unity, indicating that internal Faraday rotation in a uniform slab permeated by a regular magnetic field is not responsible for the observed jet RM. In addition, we frequently measure EVPA rotations larger than $45^\circ$ with good $\lambda^2$ fits at various locations in the jet at all frequencies as shown in Figure~\ref{extendedRM}. The fact that we could not find any statistically significant difference in RMs obtained at different frequencies at a given distance also supports an external origin (Figure~\ref{radial}). Thus, the systematic decrease of RM shown in Figure~\ref{radial} must originate from the magnetized plasma outside the jet (external Faraday rotation).

\begin{figure*}[!t]
\begin{center}
\includegraphics[width = \textwidth]{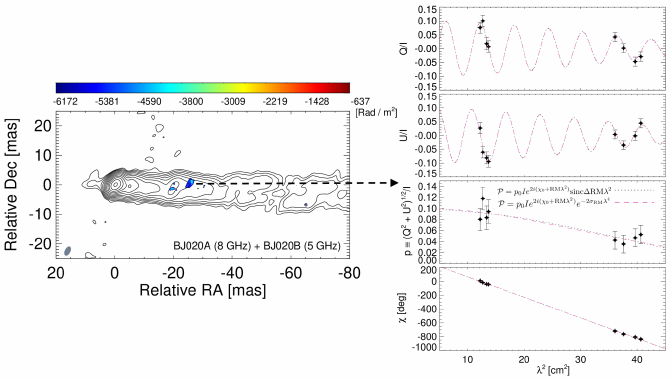}
\caption{\emph{Left:} an RM map obtained by combining BJ020A and BJ020B data sets. Contours start at 0.54 mJy per beam and increase by factors of 2. \emph{Right:} $Q/I$, $U/I$, $p$, and $\chi$ as functions of $\lambda^2$ from top to bottom. The black dotted line and the red dashed line are the best-fit of model 1 ($\sigma_{\rm RM} = 0$ in Equation~\ref{depol}) and model 2 ($\Delta \rm RM = 0$) to the data points, respectively. (see Section~\ref{sectinternal} for more details). \label{bj020}}
\end{center}
\end{figure*}

However, internal Faraday rotation might still be responsible for depolarization. As already noted in Section~\ref{sectmaps}, at many locations of the jet linearly polarized emission is not detected in our data, making the RM distributions patchy. In general, depolarization originates from either internal Faraday rotation or spatial variations in the RMs of the external Faraday screen on scales smaller than the resolution of the observations (e.g., \citealt{Burn1966, Tribble1991, Sokoloff1998, Homan2012}). The depolarization mechanism of AGN jet emission has been extensively investigated recently, thanks to observations with large bandwidths (e.g., \citealt{OSullivan2012, OSullivan2017, Hovatta2018, Pasetto2018}), or VLBI observations at many different observing frequencies (e.g., \citealt{Kravchenko2017}). Investigating the depolarization mechanism of the M87 jet is difficult for us because we have a limited number of observing frequencies with relatively short $\lambda^2$ spacings available. However, we found that the data collected in the BJ020A and BJ020B sessions could be combined because their observing dates and frequencies are relatively close to each other (Table~\ref{Info}).

We obtained the RM map as described in Section~\ref{reduction} after considering a core-shift effect between 5 and 8 GHz by employing two-dimensional cross correlation of the optically thin emission regions in the image plane \citep{CG2008} and present the map in the left panel of Figure~\ref{bj020}. We note that the results are not significantly affected by the core-shift. Significant RMs were detected in small parts of the jet because linear polarization at 5 GHz has not been detected in most parts of the jet in the inner jet region (at distances less than $\approx60$ mas), where the jet emission could be detected at 8 GHz. Nevertheless, an RM patch was detected at $\approx25$ mas from the core over a region with a size comparable to the beam size. In the right panel of Figure~\ref{bj020}, we present $Q/I$, $U/I$, $p \equiv \sqrt{Q^2+U^2}/I$, and $\chi$ as a function of $\lambda^2$ in this region.

In order to investigate the depolarization mechanism, we tried to model the Stokes I, Q, and U intensity simultaneously at different wavelengths, known as the qu-fitting technique (e.g., \citealt{Farnsworth2011, OSullivan2012}). We used a model for the complex polarization which includes the effect of depolarization due to random magnetic fields ($\sigma_{\rm RM}$) and ordered magnetic fields ($\Delta \rm RM$), given by
\begin{equation}
\mathcal{P} = p_0Ie^{2i(\chi_0+{\rm RM}\lambda^2)}e^{-2\sigma^2_{\rm RM}\lambda^4}{\rm sinc\Delta RM}\lambda^2,
\label{depol}
\end{equation}
\citep{Sokoloff1998}. We followed a recent study which detected a very high rotation measure of $(3.6\pm0.3)\times10^5\rm\ rad/m^2$ in the quasar 3C 273 with Atacama Large Millimeter Array (ALMA) observations at 1 mm \citep{Hovatta2018} and fitted Equation~\ref{depol} with $\sigma_{\rm RM} = 0$ (model 1, the black dotted lines in the right panel of Figure~\ref{bj020}) and with $\Delta \rm RM = 0$ (model 2, the red dashed lines) to the data points. The best-fit parameters are ($p_0 = 0.10\pm0.01$, $\Delta \rm RM = 532\pm62\ rad/m^2$, $\chi_0 = 184\pm6^\circ$, ${\rm RM} = -5195\pm43\rm\ rad/m^2$) and ($p_0 = 0.10\pm0.01$, $\sigma_{\rm RM} = 171\pm25\rm\ rad/m^2$, $\chi_0 = 184\pm6^\circ$, ${\rm RM} = -5194\pm43\rm\ rad/m^2$) for model 1 and 2, respectively. Both models can explain the data well with the reduced chi-square $\chi_r^2 \equiv \chi^2 / {\rm d.o.f}$, where $\rm d.o.f$ is the degree of freedom, of 0.66 and 0.64 for model 1 and 2, respectively. This is due to the sparse sampling of the data in the $\lambda^2$ space, which prevented us from solving the degenaracy. Nonetheless, the observed depolarization at $\approx 25$ mas from the core is likely due to a gradient in RM by $\approx 532 \rm \ rad/m^2$ either in the jet or in the external Faraday screen across the beam or due to random magnetic fields with $\sigma_{\rm RM} \approx 171\rm \ rad/m^2$ in the external screen \citep{Sokoloff1998, OSullivan2017, Hovatta2018, Pasetto2018}. We also obtained good $\lambda^2$ fits for the EVPA rotation larger than $4\pi$, supporting an external origin of the observed RM. The observed RM of $\approx-5194\pm43\rm\ rad/m^2$ for model 2 is consistent with that obtained in the same location by using only BJ020A (8 GHz) data, $-5535\pm1226\rm\ rad/m^2$, within $1\sigma$ and BJ020B (5 GHz) data, $-4469\pm431\rm\ rad/m^2$, within less than $2\sigma$. The deviation larger than $1\sigma$ in the latter case might be due to a non-negligible time gap of $\approx18$ days between the two data sets.

\begin{figure*}[!t]
\begin{center}
\includegraphics[trim=0mm 0mm 0mm 0mm, clip, width = 175mm]{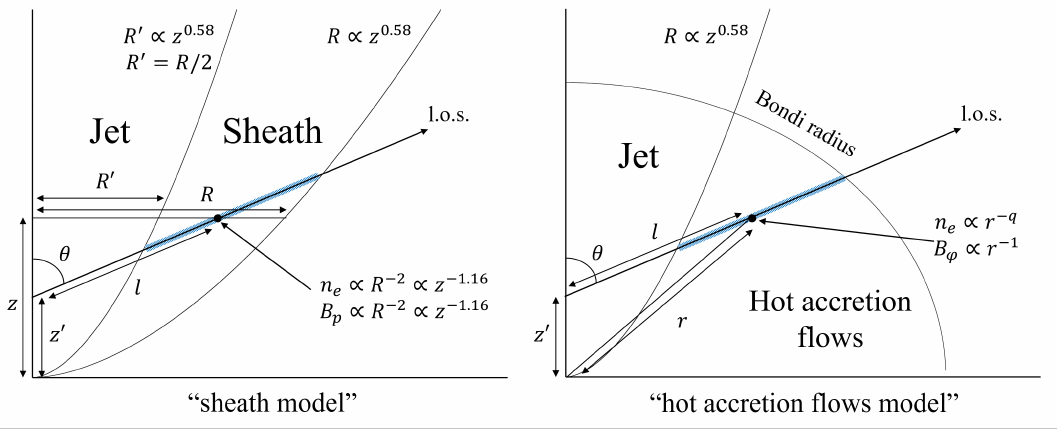}
\caption{\emph{Left:} Schematic diagram showing the sheath model. $l$ denotes the path from the emitter to the observer at the jet distance $z'$. $\theta$ is the jet viewing angle and R is the radius of the sheath. We performed numerical integration of ${\rm RM} \propto \int n_e(l)B(l)dl$ in the sheath region along each line of sight for each $z'$ (the blue shaded line), with the scaling relations of $n_e(z) \propto R^{-2} \propto z^{-1.16}$ and $B_p(z) \propto R^{-2} \propto z^{-1.16}$ provided (see Section~\ref{sectsheath} for details). \emph{Right:} Same as the left panel but showing the case of the hot accretion flows model. $r$ denotes the spherical radius and $n_e(r) \propto r^{-q}$ and $B(r) \propto r^{-1}$ are used for numerical integration of ${\rm RM} = 8.1\times10^5\int n_e(l)B(l)dl$ for each jet distance (see Section~\ref{secthot} for details). We note that the jet radius and $\theta$ are much smaller than those shown in these diagrams in reality. \label{diagram}}
\end{center}
\end{figure*}

\subsubsection{A jet sheath}
\label{sectsheath}

If the Faraday screen is placed in the immediate vicinity of the jet, e.g., like a sheath surrounding the jet as claimed for other distant AGNs (e.g., \citealt{ZT2004, Jorstad2007, OG2009, Hovatta2012, Park2018}), then one expects significant RM gradients across the jet with a possible change of the sign of the RM; this is seen in numerical simulations \citep{BM2010}. This signature has indeed been frequently observed in the jets of many blazars (e.g., \citealt{Asada2002, Asada2008, Gabuzda2004, Gabuzda2015, Gabuzda2018, Hovatta2012}). The transverse RM gradients are related to toroidal magnetic fields in the jet and/or in the sheath, which can be naturally produced in the inner part of the accretion disk and/or in the black hole's ergosphere. These magnetic fields play a crucial role in launching and powering of relativistic jets \citep{Meier2012}. MHD theories predict that poloidal magnetic fields which are dominant near the jet base become rapidly weak at larger distance and toroidal fields become dominant relatively far from the black hole (e.g., \citealt{VK2004, Komissarov2009}).

However, for M87 the observed sign of RM is negative almost everywhere inside the Bondi radius (Figure~\ref{radial}). Furthermore, we found that there is no significant difference between the RMs on the northern and southern jet edges at a given distance and the RMs appear to vary only as function of radial distance (see Appendix~\ref{sectridge}). Recently, linear polarization structure of the core of the M87 jet at 43 GHz has been revealed, showing the inferred magnetic field vectors wrapped around the core\footnote{We note that we could not obtain intrinsic (RM-corrected) EVPAs with our data sets because the data are sampled in limited wavelength ranges relatively far from $\lambda = 0$. This leads to very large uncertainties in the intrinsic EVPAs usually larger than $90^\circ$ .} \citep{Walker2018}. This suggests that toroidal fields might be dominant already on scales of $\approx 100\ r_{\rm s}$, which makes it difficult to explain the observed single (negative) RM sign and no significant difference in RMs between the north and south edges with the Faraday screen consisting of a jet sheath.


We checked whether the observed RMs can be explained by the sheath model or not if poloidal magnetic fields are somehow dominant in the sheath at distance $\gtrsim$ 5,000 $r_{\rm s}$, as indicated by a recent study of time variable RM in the radio core of a nearby BL Lac object Mrk 421 \citep{Lico2017}. We assumed (i) the same parabolic geometry of the sheath as that observed for the jet, i.e., $z \propto R^{1.73}$ \citep{AN2012, NA2013} with the radius of the outer boundary of the sheath being twice the radius of the jet (see the left panel of Figure~\ref{diagram}), (ii) a constant velocity of the sheath at different distances, (iii) no reversal in the magnetic field direction along the line of sight, and (iv) the sheath consisting of non-relativistic cold plasma. These assumptions led us to the scaling relations of $n_e(z) \propto R^{-2} \propto z^{-1.16}$ and $B_p(z) \propto R^{-2} \propto z^{-1.16}$ with $R$ being the radius of the sheath and $B_p$ the poloidal magnetic field strength. We integrated ${\rm RM} \propto \int n_e(l)B(l)dl$ numerically along each line of sight for each RM data point between the jet boundary and the sheath boundary (see the left panel of Figure~\ref{diagram}) and fitted this function to the data points at different distances with a coefficient left as a free parameter. The best-fit model is indicated by the dashed line in Figure~\ref{radial}.

\begin{deluxetable*}{ccccc}[t!]

\tablecaption{Comparison of the models}
\tablehead{
\colhead{Model} & \colhead{$n_e$ profile} &
\colhead{$B$ profile} &
\colhead{$\chi_r^2$} & \colhead{BIC} \\
(1) & (2) & (3) & (4) & (5)
}
\startdata
 Jet sheath & $n_e(z) \propto z^{-1.16}$ (fixed) & $B(z) \propto z^{-1.16}$ (fixed)& 1.73 & 86.8 \\
 Hot accretion flows& $n_e(r) \propto r^{-1.00\pm0.11}$ (fit)& $B(r) \propto r^{-1}$ (fixed)& 1.16 & 62.4
\enddata
\tablecomments{(1) Model applied to the RM data. (2) Density profile. The definition of $z$ and $r$ is explained in Figure~\ref{diagram}. (fixed) means that the fixed profile is used in the model, whereas (fit) means that the index in the power-law is left as a free parameter in the fitting. (3) Magnetic field strength profile. (4) Reduced chi-square for the best-fit. (5) Bayesian Information Criteria for the best-fit. The number of data points used in the fitting is 49. \label{tablecomp}}

\end{deluxetable*}

\subsubsection{Hot accretion flows}
\label{secthot}

We use the scaling relations $n_e(r) = n_{e, \rm out}(r/r_{\rm out})^{-q}$ with $0.5\leq q \leq 1.5$ and $B(r) = B_{\rm out}(r/r_{\rm out})^{-1}$, where $r$ is the radial distance from the black hole and $n_{e, \rm out}$ and $B_{\rm out}$ are the electron number density and the magnetic field strength at $r_{\rm out}$, respectively. The former is based on self-similar solutions of hot accretion flows \citep{BB1999, YN2014}. The latter is based on the assumption that toroidal magnetic fields are dominant in the accretion flows (e.g., \citealt{Hirose2004}). We note that we are restricted to 1D scaling relations due to the limitations of the 2D accretion flow models including non-negligible magnetic fields currently available, especially at small polar angles which is of our interest because of the small jet viewing angle (e.g., \citealt{Mosallanezhad2016, BM2018}). In other words, we assume here that the quantities of the flows would be spherically symmetric for regions with a polar angle smaller than the jet viewing angle of $17^\circ$. 

We employed ${\rm RM} = 8.1\times10^5\int n_e(l)B(l)dl$ ($\rm RM$ in units of $\rm rad/m^2$, $n_e$ in units of $\rm cm^{-3}$, $B$ in units of Gauss, and $l$ in units of parsec; \citealt{GW1966}) for `cold' non-relativistic plasma, which applies to the relatively large spatial scales probed in this study \citep{YN2014}. We also performed numerical integration along each line of sight between the jet boundary and the Bondi radius (see the right panel of Figure~\ref{diagram}, see also Section~\ref{sectoutside} for discussion of the potential contribution by gas outside the Bondi radius). The result of fitting this function to the observed RM values measured inside the Bondi radius is indicated by the solid line in Figure~\ref{radial} with the best-fit parameter of $q=1.00\pm0.11$, which indicates $\rho\propto r^{-1}$ with $\rho$ being the mass density. We could also obtain $n_{e, \rm out}B_{\rm out}$ from the fitting and when using $n_{e, \rm out} \approx 0.3\rm\ cm^{-3}$ at the Bondi radius measured by the X-ray observations \citep{Russell2015}, we obtain $B_{\rm out} = (2.8\pm0.8)\times10^{-6}\rm\ G$.

\section{Discussion}

\subsection{Jet sheath vs hot accretion flows}
\label{sectcomp}

In Section~\ref{sectscreen}, we considered three different sources of Faraday rotation, (i) the jet itself, (ii) a sheath surrounding the jet, and (iii) hot accretion flows. Given that the observed EVPA rotations are larger than $45^\circ$ at various locations in the jet with good $\lambda^2$ scalings and the observed degree of linear polarization is usually much higher than that expected in the internal Faraday rotation model, we excluded the scenario (i) in Section~\ref{sectinternal}. Although the hot accretion flows model (the solid line in Figure~\ref{radial}) apparently fits the data better than the sheath model (the dashed line in Figure~\ref{radial}), a statistical analysis is necessary to properly determine the better model. In Table~\ref{tablecomp}, we present the values of reduced chi-square ($\chi_r^2$) and Bayesian information Criterion (BIC) obtained in the best-fit for each model. The BIC is defined as ${\rm{BIC}} \equiv -2\ln \mathcal{L}_{\rm max} + k\ln N$, where $\mathcal{L}_{\rm max}$ is the maximum likelihood and $-2\ln \mathcal{L}_{\rm max}$ is equivalent to the $\chi^2$ value for the best-fit model in case for Gaussian errors (when neglecting a constant term), $k$ the number of free parameters in the model, and $N$ the number of data points used in the fit. The BIC allows one to compare the goodness of fit of different models having different numbers of free parameters \citep{Schwarz1978}. The difference between the BIC values ($\Delta \rm BIC$) for two models quantifies how strongly one model is preferred over the other one, where a model with a lower BIC value is more favored by the data. Conventionally, $0<\Delta \rm BIC<2$ represents weak evidence, $2<\Delta \rm BIC<6$ positive evidence, $6<\Delta \rm BIC<10$ strong evidence, and $10<\Delta \rm BIC$ very strong evidence (e.g., \citealt{Jeffreys1961, KR1995, Mukherjee1998, Liddle2004}). The value of BIC for the hot accretion flows model is smaller than that for the jet sheath model by $\approx24$ (Table~\ref{tablecomp}), indicating that the former is strongly favored by the data. 

\begin{figure*}[!t]
\begin{center}
\includegraphics[trim=3mm 62mm 7mm 9mm, clip, width = 179mm]{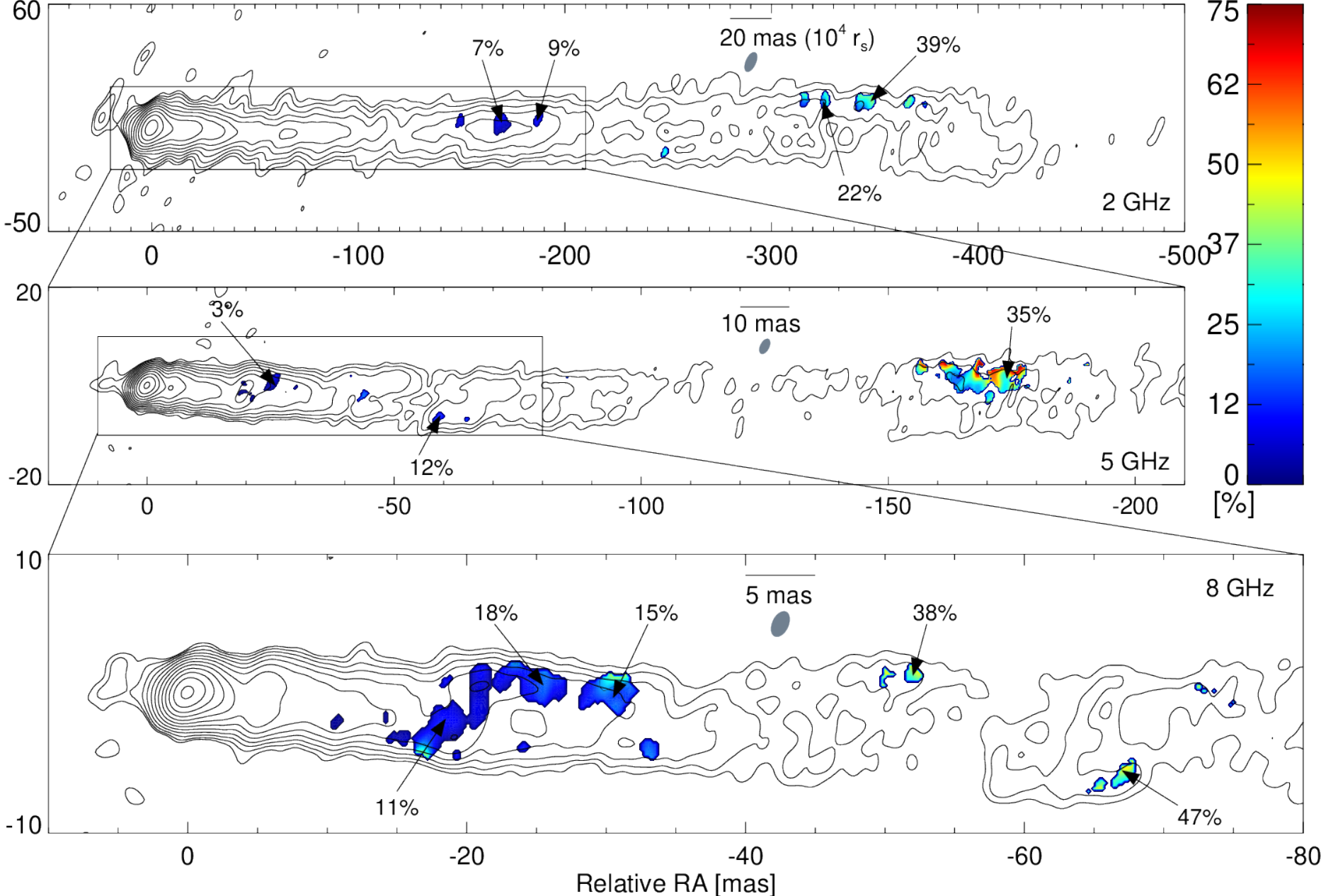}
\caption{Same as Figure~\ref{RM} but colors show degree of linear polarization for the first sub-band data at each frequency. Contours start at 0.79, 0.54, and 0.53 mJy per beam for the 2 GHz, 5 GHz, and 8 GHz maps, respectively, and increase by factors of 2. The values of fractional polarization at various locations of the jet are noted. \label{frac}}
\end{center}
\end{figure*}

We note that the above conclusion is based on the results obtained by using several assumptions on the jet sheath. For example, we assumed that the sheath geometry is the same as the jet, which may not be true. When we relax this assumption and leave the power-law index in the width profile of the sheath as a free parameter, i.e., $z_{\rm sheath}\propto R_{\rm sheath}^\eta$, and fit the sheath model to the data points, then we obtain the best-fit with $\eta = 2.49\pm0.17$. This indicates that the sheath is more strongly collimated than the jet, which is unlikely because the inner part (closer to the axis) of streamlines is thought to be more collimated than the outer part for collimated outflows (e.g., \citealt{Komissarov2007, Komissarov2009, Tchekhovskoy2008, Nakamura2018}). Or, if we assume that toroidal fields are dominant in the sheath and fix the sheath geometry, we obtain a relatively good fit with a $\rm BIC$ value comparable to that of the hot accretion flows model. However, as noted in Section~\ref{sectsheath}, it is difficult to explain the absence of a systematic difference between the RMs on the south and north edges in this case.

An alternative scenario is that the Faraday screen consists of dense clouds with ordered magnetic fields that are entrained by the jet (suggested by \citealt{ZT2002}). The volume filling factor of these clouds, if they exist, is expected to be very small and this might explain why the RMs are detected in only small parts of the jet. Although we could not exclude this possibility, the observed depolarization at longer wavelengths does not seem to support this scenario. We present the distribution of the degree of linear polarization overlaid on the contours of total intensity emission for one observation at each frequency in Figure~\ref{frac}. At higher observing frequencies, the distribution of significant linear polarization becomes more continuous and the degree of linear polarization becomes higher at a given distance, notably at $\approx20$ and $\approx 170$ mas from the core. This suggests that the Faraday screen consists of a continuous and extended medium such as winds but significant depolarization in large parts of the jet makes the observed patchy RM distributions especially at lower frequencies. We will investigate the depolarization mechanism at various locations of the jet with dedicated multi-frequency polarimetric observations in the near future, which will allow us to identify the Faraday screen more rigorously.

Taken as a whole, we conclude that attributing the Faraday screen to hot accretion flows is most consistent with the data presented in this paper and we discuss the results obtained by applying the hot accretion flows model hereafter.


\subsection{Winds and the Faraday screen}
\label{sectwinds}

The density profile we derived is significantly flatter than the profile $\rho \propto r^{-1.5}$ from the ADAF with pure gas inflows \citep{NY1994, NY1995a}, at a level of $>3\sigma$. Instead, our observations are in good agreement with the ADIOS model \citep{BB1999, YN2014}, suggesting that substantial winds from hot accretion flows exist in M87. Our results are consistent with the results of various numerical simulations of hot accretion flows, i.e., $\rho \propto r^{-q}$ with $q=0.5-1$ (see \citealt{Yuan2012b} and references therein). Since our study probes regions relatively far from the central engine, i.e., $\gtrsim$ 5,000 $r_{\rm s}$, the results of \cite{Pang2011} would be the most suitable to compare with our observations among various simulations. They performed a numerical survey with various parameters of the accretion flows in their 3D MHD simulations, in which the outer boundary is extended up to ten times the Bondi radius, and found the most favored value of $q\approx1$. This result is in good agreement with our finding. We note that previous observations of Faraday rotation at 1 mm with the Submillimeter Array already ruled out the pure inflow scenario \citep{Kuo2014}, which is consistent with our results. However, we could further constrain the accretion model of M87 from the radial RM profile measured at distances over nearly two orders of magnitude.

GRMHD simulations also found the production of winds, which are non-relativistic, moderately magnetized gas outflows surrounding the highly magnetized and collimated jets\footnote{The geometry of winds is approximated as conical \citep{Sadowski2013, Yuan2015} and the use of $B_\phi \propto r^{-1}$ in our modelling (Section~\ref{secthot}) would be valid because $B_\phi \propto R^{-1} \propto r^{-1}$.} (e.g., \citealt{Sadowski2013, Nakamura2018}). Since the viewing angle of the M87 jet is relatively small ($\theta \approx 17^{\circ}$ \citealt{Mertens2016}), it is reasonable to regard the winds as a dominant source of the observed RMs and thus as an external medium confining the jet. Nevertheless, we note that the contribution of weakly magnetized inflows to the observed RMs is probably non-negligible. From the derived pressure profile and the assumed magnetic field configuration for winds, one expects $\beta \equiv p_{\rm gas}/p_{\rm mag} \approx 68$ at $r_{\rm out}$ assuming $\beta\approx1$ close to the black hole \citep{DeVilliers2005} because $p_{\rm gas} \propto r^{-5/3}$ and $p_{\rm mag} \propto r^{-2}$ with $p_{\rm mag}$ being the magnetic pressure (see Section~\ref{sectcoll}). However, we obtained $\beta \approx 1400$ at $r_{\rm out}$ using $B_{\rm out}\approx2.8\ \mu G$ from the fitting (Section~\ref{secthot}) and the pressure at $r_{\rm out}$ measured by X-ray observations \citep{Russell2015}. This $\beta$ is larger than that for winds by an order of magnitude and we expect some contribution of weakly magnetized inflows to the observed RMs \citep{YN2014}. Thus, the Faraday screen of the M87 jet might consist of a complex mixture of inflows and winds.

\subsection{Jet collimation by winds}
\label{sectcoll}

The pressure profile of an external medium surrounding the jet can be estimated from the density profile. Assuming an adiabatic equation of state for non-relativistic monatomic gas, the pressure scales like $p_{\rm gas} \propto \rho^\gamma \propto r^{-5/3}$, where $\gamma=5/3$ is the specific heat ratio. According to MHD models, AGN jets are gradually accelerated by transferring the electromagnetic energy of the flow to its kinetic energy (e.g., \citealt{Komissarov2009, Lyubarsky2009, TT2013}). Jet collimation is critical for the conversion; therefore the acceleration and collimation zones in AGN jets are expected to be co-spatial \citep{Marscher2008}. It has been shown that the flow acceleration is very inefficient without an external confinement (e.g., \citealt{Eichler1993, BL1994}). If the pressure profile of the external medium follows a power-law, i.e., $p_{\rm ext} \propto r^{-\alpha}$, the power-law index must satisfy $\alpha \leq 2$ to permit for a parabolic jet shape \citep{BL1994, Lyubarsky2009, Komissarov2009, Vlahakis2015}. Our results, $\alpha = 1.67$ for the external medium, and the observed parabolic geometry up to the Bondi radius \citep{Junor1999, AN2012, NA2013, Hada2013}, are consistent with the MHD collimation-acceleration scenario\footnote{We note, however, that $\alpha = 1.67$ leads to an asymptotic jet shape with $z \propto R^{2.4}$ in the MHD models \citep{Lyubarsky2009}, which deviates from the observed one, $z \propto R^{1.73}$ \citep{NA2013}. Also, the fact that the jet appears stable over a large distance range can be explained by the loss of causual connectivity across the jet, if $\alpha>2$ \citep{PK2015}. However, the jet becomes conical in this case. We note that if the same temperature profile as in the ADAF self-similar solutions, $T\propto r^{-1}$, can be applied to the ADIOS model \citep{Yuan2012b}, then we obtain $\alpha = 2$ which allows $1<a<2$ in $z\propto R^a$ \citep{Komissarov2009}. However, this requires a remarkable coincidence, considering the non-negligible error in the obtained density profile $\rho \propto r^{-1.00\pm0.11}$. Therefore, we adopt $\alpha = 1.67$ obtained from the assumption of a simple equation of state, which generally allows a parabolic jet geometry (see Section 5 in \citealt{PK2015} for more discussion).} \citep{Komissarov2009, Lyubarsky2009, Vlahakis2015}. Indeed, systematic acceleration of the M87 jet inside the Bondi radius has been discovered \citep{Asada2014, Mertens2016, Hada2017, Walker2018}. Remarkably, recent GRMHD simulations presented that non-relativistic winds launched from hot accretion flows play a dynamical role in jet collimation and the jet is accelerated to relativistic speeds \citep{Nakamura2018}. We note that our conclusion is also supported by the fact that the observed collimation profile of the M87 jet was successfully modelled by a two-zone MHD model, where an inner relativistic jet is surrounded by highly magnetized \citep{Gracia2005, Gracia2009} or weakly magnetized \citep{GL2016} non-relativistic outer disk winds. We also note that the confinement of the jet by hot accretion flows and/or winds on smaller scales has been suggested by \cite{Hada2016}, where a complicated innermost collimation profile with a local constricted jet structure was observed.

\subsection{Mis-alignment}
\label{sectmis}

\begin{figure}[!t]
\begin{center}
\includegraphics[trim=0mm 0mm 0mm 0mm, clip, width = 88mm]{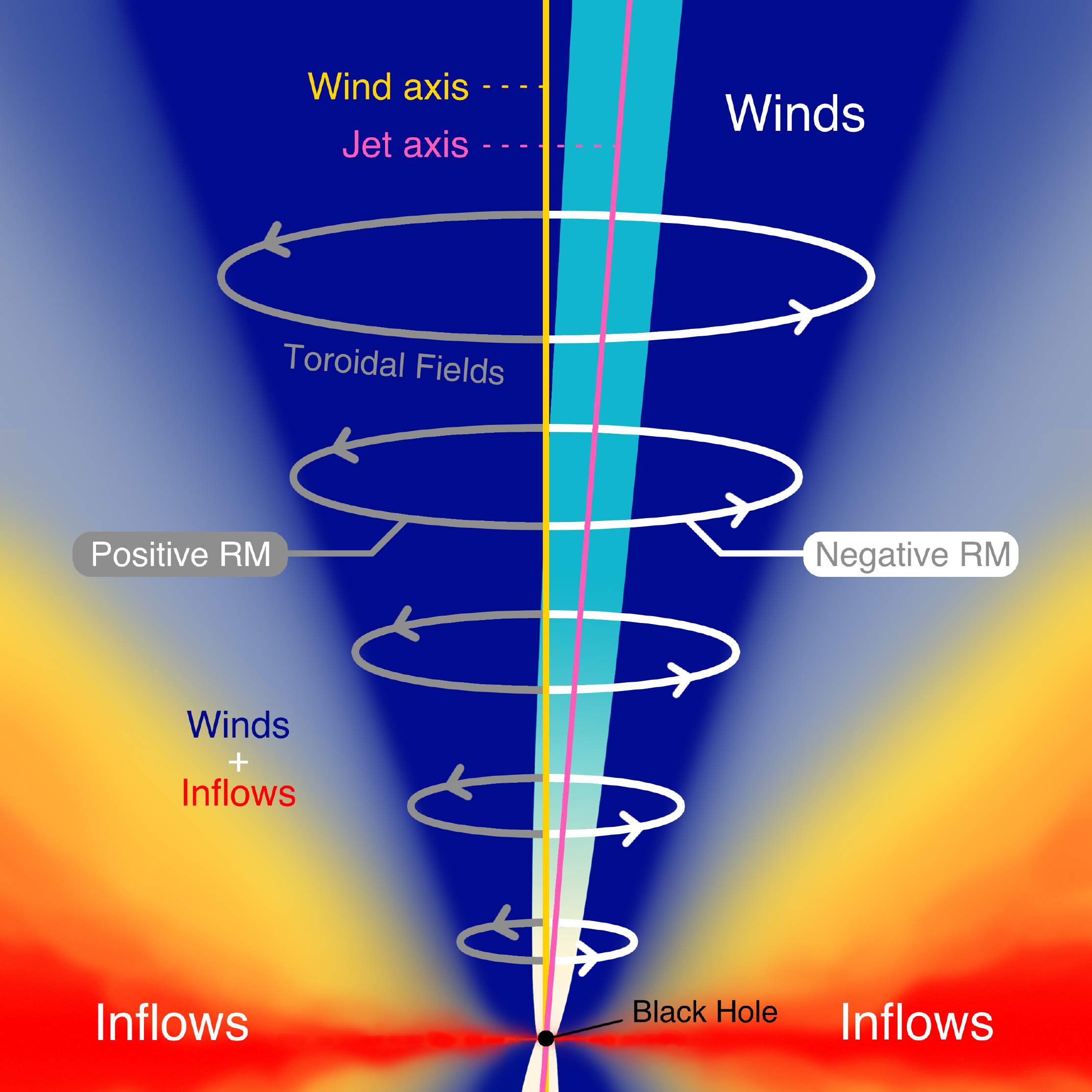}
\caption{Schematic diagram of the black hole inflow--outflow system in M87. Different colors represent regions dominated by dense, hot, and turbulent inflows (red and yellow), collimated and highly magnetized jets (cyan), non-relativistic and moderately magnetized winds (dark blue), and a complex mixture of inflows and winds (light blue). The winds are permeated by toroidal magnetic fields indicated by gray and white loops. The jet axis (purple vertical line) is tilted with respect to the wind axis (yellow vertical line) and the jet exposes only one side of the toroidal fields, resulting in a single (negative) RM sign from the point of view of a distant observer. \label{schematic}}
\end{center}
\end{figure}

The dominance of a single RM sign for M87 implies that the background light source, i.e., the jet, exposes only one side of the toroidal magnetic loops in the Faraday screen. This situation can be realized when there is a mis-alignment between the jet axis and the symmetry axis of the toroidal field loops (Figure~\ref{schematic}). This is another indication for winds or inflows as the dominant source of Faraday rotation because the jet sheath is tightly attached to the jet and cannot be tilted relative to the jet axis. Since the jet is highly collimated and narrow \citep{Junor1999, AN2012, Doeleman2012}, only a slight misalignment by $\approx5^\circ$ can result in observations of a fixed RM sign over a large distance range. Such small misalignments seem to be quite common in hot accretion flows even when the magneto-spin alignment effect, an alignment of the accretion disk and jets with the black hole spin by strong magnetic fields near the black hole, operates \citep{McKinney2013}.

We note that it is unlikely that poloidal magnetic fields are responsible for the observed RMs of M87 because in that case one expects $\rho \propto r^0$ from $B \propto r^{-2}$, which is impossible to explain with the accretion models currently available \citep{YN2014}. However, there is indication of non-negligible poloidal fields as well as toroidal fields -- resulting in helical magnetic fields -- in the jet environment of other distant AGNs, which results in transverse RM gradients with no sign changes (e.g., \citealt{Asada2002, Zamaninasab2013, Gomez2016, Gabuzda2018}, see also Section~\ref{sectsheath}). The existence of non-negligible poloidal fields was indicated even for the M87 jet at HST-1 from the observed moving knots with both fast and slow velocities which could be explained by quad relativistic MHD shocks in a helical magnetic field permeating the jet \citep{Nakamura2010, NM2014}. In Section~\ref{sectsheath} and Section~\ref{sectcomp}, we explained that poloidal magnetic fields might be very weak at distances $\gtrsim$ 5,000 $r_{\rm s}$ probed in this study and we concluded that hot accretion flows and winds are more probable to be the Faraday screen than the jet sheath. However, if the jet experiences recollimation, which may lead to formation of standing shocks (e.g., \citealt{DM1988, Gomez1995, Agudo2001, Mizuno2015, Marti2016, Fuentes2018}), then the strength of poloidal fields could be substantially enhanced. Indeed, the width of HST-1 is significantly smaller than expected from the parabolic (conical) width profile inside (outside) the Bondi radius \citep{AN2012}, which has been explained with a hydrodynamic recollimation shock (e.g., \citealt{Stawarz2006, BL2009, AN2012}). Also, the core of blazars is often identified with a recollimation shock (e.g., \citealt{DM1988, Marscher2008a, Cawthorne2013}). This may explain the presence of non-negligible poloidal fields in the sheath of blazar jets and in HST-1, but not in the M87 jet inside the Bondi radius.

\subsection{Mass accretion rate}
\label{sectrate}

The presence of winds indicates that the actual rate of mass accreted onto the black hole could be substantially smaller than the Bondi accretion rate. If the density profile in the equatorial plane is similar to the one we observe, i.e., if a radial self-similarity holds, one expects $\dot{M}(r) = \dot{M}_{\rm ADAF}(r/r_{\rm out})^{1.5-q}$ (e.g., \citealt{BB2004, Yuan2012b, YN2014}), where $\dot{M}_{\rm ADAF}$ is the mass accretion rate in the classical ADAF model. Using $\dot{M}_{\rm Bondi} = 0.1M_\odot\rm yr^{-1}$ \citep{Russell2015} and $\dot{M}_{\rm ADAF}=0.3\dot{M}_{\rm Bondi}$ with a viscosity parameter $\alpha=0.1$ \citep{NF2011}, assuming a constant mass accretion rate inside $10\ r_{\rm s}$ \citep{Yuan2012b}, the rate of mass passing through the event horizon of M87 would be $\dot{M}_{\rm BH} \approx 0.3\dot{M}_{\rm Bondi}(10r_{\rm s}/3.6\times10^5r_{\rm s})^{0.5} = 1.6\times10^{-4}M_\odot\rm yr^{-1}$. 

This is consistent with the upper limit on the accretion rate of $9.2\times10^{-4}M_\odot\rm yr^{-1}$ obtained from previous polarimetric observations of M87 at 1 mm \citep{Kuo2014}. We obtained a radiative efficiency $\epsilon\equiv L_{\rm disk} / \dot{M}_{\rm BH}c^2 \approx 3.8\%$ for a disk luminosity of $L_{\rm disk} = 3.4\times10^{41}\rm\ erg\ s^{-1}$ \citep{Prieto2016} and $\dot{M}_{\rm BH}/\dot{M}_{\rm Edd} \approx 1.2\times10^{-6}$, where $\dot{M}_{\rm Edd}\equiv 10L_{\rm Edd}/c^2$ with $L_{\rm Edd}$ being the Eddington luminosity \citep{YN2014}. This is consistent with recent theoretical studies which found that the radiative efficiency of hot accretion flows might not be as small as previously thought even at very low accretion rates \citep{XY2012, YN2014}. The obtained radiative efficiency is consistent with the case of $\delta=0.5$ in \cite{XY2012}, where $\delta$ is the fraction of the viscously dissipated energy in the accretion flows used to directly heat electrons. Remarkably, this is similar to the value found for Sgr A* in the SED modelling \citep{Yuan2003}. Our results indicate that a very low accretion rate due to the mass loss via winds is probably the main reason for the faintness of the active nucleus of M87 and a similar conclusion was drawn for Sgr A* from the measured RMs \citep{Bower2003}. 

The accretion rate we derive suggests a jet production efficiency of $\eta\equiv P_{\rm jet} / \dot{M}_{\rm BH}c^2 \gtrsim 110\%$ with a jet power $P_{\rm jet} \gtrsim 10^{43} \rm\ erg\ s^{-1}$ for M87 (e.g., \citealt{BB1996, Owen2000, Allen2006, Rafferty2006, Stawarz2006, BL2009}, see \citealt{Broderick2015} for more discussion). This is higher than the efficiency of gravitational binding energy of accretion flows released as radiation in a maximally rotating black hole by a factor of three \citep{Thorne1974} and indicates that almost all of input rest mass power is released as jet power. This is possible only when (i) the accretion disk of M87 is in magnetically arrested disk (MAD) state in which the magnetic pressure of the poloidal magnetic fields is balanced by the ram pressure of the accreting gas \citep{Narayan2003, Tchekhovskoy2011, McKinney2012} and (ii) there is extraction of rotational energy of a spinning black hole that powers the jet, the Blandford-Znajek (BZ) process \citep{BZ1977}. GRMHD simulations find that the efficiency of winds launched from hot accretion flows or of jets launched not in a MAD state is $\lesssim 10\%$ \citep{Sadowski2013} but can go up to $\approx 300\%$ with the BZ process in a MAD state \citep{Tchekhovskoy2011, Tchekhovskoy2012, McKinney2012, Sadowski2013}. This is also in agreement with recent observational evidence that most radio-loud active galaxies, including M87, are in a MAD state \citep{Zamaninasab2014}. The jet power larger than or comparable to the accretion power $\dot{M}_{\rm BH} c^2$ has also been found for many blazars \citep{Ghisellini2014}.

We note that the estimation of mass accretion rate and the related quantities above is based on an assumption that the gas contents of the accretion flows are dominated by hot gas. However, a recent study showed that significant amounts of cold and chaotic gas can form near or inside the Bondi radius via non-linear growth of thermal instabilities, resulting in the accretion rate being boosted up to two orders of magnitude compared to the case of hot gas only \citep{Gaspari2013}. However, as already noted in \cite{NT2015}, the amount of cold gas is unlikely to be much larger than the amount of hot gas in the accretion flows because of (i) no correlation between the jet power and the total mass of cold molecular gas in many radio galaxies \citep{McNamara2011} and (ii) not very tight but significant correlation between the jet power and the Bondi accretion power of nearby radio galaxies (e.g., \citealt{Allen2006, Balmaverde2008, Russell2013, NT2015}). In addition, even if the true accretion rate is an order of magnitude larger than the one we estimated due to the cold gas, the jet production efficiency would be still very large, possibly close to $\approx100\%$. The jet power of $\approx10^{43}\rm\ erg\ s^{-1}$ we used above is estimated from observations of X-ray cavities, which represents the mechanical power of the jet averaged over the cavity buoyance time of about $\gtrsim1$ Myr \citep{Broderick2015}. Also, this power should be in general regarded as a lower limit on the total mechanical power of the jet due to possibly missing cavities and the significant contribution of weak shocks and sound waves to the jet power, which was not considered in the cavity analysis \citep{Russell2013}. Other estimates of the jet power which reflect more recent ($\lesssim$ a few $\times 10^3$ yr) jet activities of M87 provide $\approx10^{44}\rm\ erg\ s^{-1}$ (e.g., \citealt{BB1996, Owen2000, Stawarz2006, BL2009, Broderick2015}). This may compensate for the increased mass accretion rate due to cold gas and a high jet production efficiency would still be maintained.

The magnetic flux near the event horizon in a MAD state is saturated at $\sim50\left(\dot{M}_{\rm BH}r^2_{\rm g}c\right)^{1/2}\rm G\ cm^{2}$, where $r_{\rm g}\equiv GM_{\rm BH}/c^2$ is the black hole gravitational radius, $G$ the gravitational constant \citep{Tchekhovskoy2011}. One can estimate the magnetic field strength at the horizon via $B_{\rm MAD} \approx \Phi_{\rm MAD} / 2\pi r^2_{\rm g} = 10^{10}(M/M_\odot)^{-1/2}(\dot{M}_{\rm BH}/\dot{M}_{\rm Edd})^{1/2}\rm\ G$  \citep{YN2014}. We obtain $B_{\rm MAD} \approx 142 \rm\ G$, which is roughly consistent with the magnetic field strength limit provided by \cite{Kino2015}, $50\lesssim B_{\rm tot} \lesssim 124 \rm\ G$, in the presence of an optically thick region with synchrotron self-absorption near the jet base. This indicates that the jet base might be highly magnetized and the jet can be accelerated by the Poynting flux conversion \citep{McKinney2006, Komissarov2007, Komissarov2009, Lyubarsky2009}.

\subsection{RM at HST-1}
\label{sectHST}

The sudden increase of RM at HST-1 by a factor of $\approx10$ compared to those values at $\approx2\times10^5\ r_{\rm s}$ with positive RM sign may require explanations that are different from the case of RMs inside the Bondi radius. This is because HST-1 is located outside the Bondi radius and thus the contribution of inflows and outflows to the observed RMs is probably small. A simple explanation would be a compact gas cloud located in the line of sight toward HST-1 with very high electron density and/or magnetic field strengths, which might be the case for a nearby radio galaxy 3C 84 \citep{Nagai2017}. However, this requires a remarkable coincidence because most of the jet region on relatively large spatial scales observed with the VLA show much smaller RMs well represented by $\approx130\rm\ rad/m^2$ \citep{Algaba2016}. We could not observe any significant jump in RM at a specific distance from the black hole in the inner jet region and it is unlikely that a compact cloud with high Faraday depth is located only in the line of sight toward HST-1.

Another possible explanation is a recollimation shock which has been proposed to explain the compactness of HST-1 and its temporal variability (e.g., \citealt{Stawarz2006, BL2009}, see Section~\ref{sectmis} for more details). Emission from the shock is expected to concentrate near the jet axis where the pressure of the shocked gas is very high, surrounded by a relatively low-pressure region \citep{BT2018}. In this scenario, the emitting region would be quite compact and the dominant source of Faraday rotation would be the surrounding shocked jet region. This is consistent with (i) our finding that external Faraday rotation is dominant also in HST-1 and (ii) the large RM values in HST-1 which could be explained by the enhancement of thermal electron density and strong magnetic fields in the shock, on the order of mG \citep{Harris2003, Harris2009, Giroletti2012}. We will investigate the origin of the enhanced RM at HST-1 more deeply with more data sets in a forthcoming paper (Park et al. 2018, in prep.).

\subsection{EHT observations}
\label{sectEHT}

Our results indicate the presence of winds on relatively large spatial scales of $\gtrsim$ 5,000 $r_{\rm s}$. The observed continuous jet collimation profile from the vicinity of the jet base to the distance of $\lesssim$ 200,000 $r_{\rm s}$ \citep{Junor1999, Doeleman2012, AN2012, Hada2013} implies that a similar mechanism of jet collimation by the winds may be at work on smaller scales as well. On-going and future full-polarimetric observations with the EHT (e.g., \citealt{Doeleman2008, Doeleman2012, Lu2013, Akiyama2015, Johnson2015, Johnson2018, Fish2016, Lu2018}) in conjunction with the phased-up ALMA at 230 and 345 GHz will provide an unprecedented view of polarization and RM structures in the jet on scales down to a few $r_{\rm s}$ together with an image of the black hole shadow (e.g., \citealt{BL09, Dexter2012, Lu2014, Chael2016, Moscibrodzka2016, Moscibrodzka2017, Akiyama2017, Pu2017}), enabling a definitive test for the origin of winds and the jet.

\section{Conclusions}

We studied Faraday rotation in the jet of M87 with eight VLBA data sets. We found that the magnitude of RM systematically decreases with increasing distance from the black hole from 5,000 to 200,000 $r_{\rm s}$. Our work leads us to the following principal conclusions:

\begin{enumerate} 
\item We found that the degree of linear polarization in the jet is usually much higher than that expected in the case of internal Faraday rotation in a uniform slab with regular magnetic fields. In addition, we found that EVPA rotations are larger than 45$^\circ$ at various locations in the jet and always follow $\lambda^2$ scalings, which is difficult to reproduce with internal Faraday rotation in a synchrotron emitting region with a realistic geometry and magnetic field structure. We conclude that the systematic decrease of RM must originate from the magnetized plasma outside the jet, supporting an external Faraday rotation scenario.

%
\item We found that the observed sign of RM is predominantly negative inside the Bondi radius, without indication of significant difference in RMs detected on the north and south edges. The observed radial RM profile is difficult to explain with a sheath surrounding the jet permeated by poloidal magnetic fields being the Faraday screen. This implies that the Faraday screen consists of hot accretion flows, not of the jet sheath.
\item We applied hot accretion flows model to the RM data points and obtained a best-fit function consistent with $\rho\propto r^{-1}$. This result is in good agreement with the ADIOS model in which substantial winds, non-relativistic un-collimated gas outflows, are launched from hot accretion flows. The winds are likely surrounding the highly collimated relativistic jet and probably a dominant source of the observed RMs (Figure~\ref{schematic}). However, we see indication for non-negligible contribution of inflows to the observed RMs as well.
\item The density profile we obtained leads to the pressure profile of the winds, an external medium surrounding the jet, which is $p_{\rm gas} \propto r^{-5/3}$. This profile is consistent with a scenario in which the jet is substantially collimated by the winds, resulting in gradual acceleration of the jet in an MHD process. This is in agreement with the observed gradual collimation and acceleration of the jet inside the Bondi radius.
\item The negative RM sign preferentially found inside the Bondi radius indicates that the jet exposes only one side of the toroidal magnetic loops in the Faraday screen (Figure~\ref{schematic}). We conclude that the jet axis and the wind axis are mis-aligned with respect to each other. Since the jet is narrow, a slight mis-alignment by only $\approx5^\circ$ can lead to a fixed RM sign at distances $\gtrsim$ 5,000 $r_{\rm s}$. According to recent GRMHD simulations \citep{McKinney2013}, such a (small) mis-alignment seems to be common in hot accretion flows, depending on the history of gas accretion, even when the magneto-spin alignment effect operates.
\item The mass accretion rate can be substantially lower than the Bondi accretion rate due to the winds; we obtained $\dot{M}_{\rm BH} = 1.6\times10^{-4}M_{\odot}\rm yr^{-1}$, assuming a radial self-similarity of the density profile. This leads to a radiative efficiency of 3.8\% at $\dot{M}_{\rm BH}/\dot{M}_{\rm Edd} = 1.2\times10^{-6}$, which indicates that the radiative efficiency is not as small as usually assumed and the faintness of the nucleus of M87 is mainly due to the reduced mass accretion rate. Also, we obtained a jet production efficiency of $\gtrsim110$\%, implying that extraction of rotational energy of a spinning black hole might be at work in a MAD state.
\item The rotation measure at HST-1, located outside the Bondi radius, is larger by an order of magnitude and shows the opposite sign compared to the RM profile inside the Bondi radius. We conclude that this might be related with a recollimation shock that possibly forms in HST-1.
%

\end{enumerate}

We conclude with several caveats that need to be addressed in future studies. We used simple one-dimensional self-similar solutions for the density and magnetic field strength in the hot accretion flows model, while it is unclear whether this is valid or not. Studies of two dimensional solutions of hot accretion flows showed a breakdown of spherical symmetry (e.g., \citealt{Mosallanezhad2016, BM2018}), though the behavior of physical parameters measured close to the jet axis is poorly constrained yet. We assumed $\approx130\rm\ rad/m^2$ for the contribution of the diffuse gas in M87 outside the Bondi radius based on the results of RM studies of the large scale jet but this could be uncertain. We assumed that the same radial density profile holds for the polar region and for the equatorial region to estimate the mass accretion rate. This may not be true as seen in a recent study by \cite{Russell2018}, though their results are obtained relatively close to the Bondi radius. We conclude that a sheath surrounding the jet is unlikely to be the Faraday screen based on the fact that the RMs detected on the southern and northern sides of the jet at a given distance are similar to each other. However, we could not test whether there are significant transverse RM gradients in the jet due to limited sensitivity and/or substantial depolarization. We plan to perform polarimetric observations with high sensitivity and having both short and long $\lambda^2$ spacings to constrain the origin of Faraday rotation more robustly and to investigate the depolarization mechanism in the near future.

\acknowledgments 
We thank the referee for constructive comments, which helped to improve the paper. The Very Long Baseline Array is an instrument of the Long Baseline Observatory. The Long Baseline Observatory is a facility of the National Science Foundation operated by Associated Universities, Inc. This research has made use of data from the University of Michigan Radio Astronomy Observatory which has been supported by the University of Michigan and by a series of grants from the National Science Foundation, most recently AST-0607523. We acknowledge financial support from the Korean National Research Foundation (NRF) via Global Ph.D. Fellowship Grant 2014H1A2A1018695 (J.P.) and Basic Research Grant NRF-2015R1D1A1A01056807 (S.T.). M.K. acknowledges the financial support of JSPS KAKENHI program with the grant numbers of JP18K03656 and JP18H03721.

\begin{appendix}

\section{Errors in linear polarization quantities}
\label{appendixa}

In this appendix, we present the details of error estimation for linear polarization quantities. We used the following equations to estimate the errors in linear polarization quantities \citep{Roberts1994, Hovatta2012},

\begin{equation}
\sigma_P = \frac{\sigma_Q + \sigma_U}{2}
\label{polerr}
\end{equation}

\begin{equation}
\sigma_{\rm EVPA} = \frac{\sigma_P}{2P},
\label{evpaerr}
\end{equation}

%

\noindent where $\sigma_P$, $\sigma_{\rm EVPA}$, $\sigma_Q$, and $\sigma_U$ are uncertainties in the polarized intensity, EVPA, Stokes Q and U data, respectively. $\sigma_Q$ and $\sigma_U$ are estimated by adding different noise terms in quadrature, i.e.,

\begin{equation}
\sigma^2 = \sigma_{\rm rms}^2 + \sigma_{\rm Dterm}^2 + \sigma_{\rm CLEAN}^2
\label{fullerr}
\end{equation}

\begin{equation}
\sigma_{\rm Dterm} = \frac{\sigma_{\Delta}}{(N_{\rm ant} N_{\rm IF} N_{\rm scan})^{1/2}}(I^2 + (0.3I_{\rm peak})^2)^{1/2}
\label{dtermerr}
\end{equation}

\begin{equation}
\sigma_{\rm CLEAN} = 1.5\sigma_{\rm rms},
\end{equation}

\noindent where $\sigma_{\rm rms}$, $\sigma_{\rm Dterm}$, and $\sigma_{\rm CLEAN}$ denote rms noise, D-term errors, and CLEAN errors, respectively. We estimated the rms noise in the residual maps after the CLEAN procedure in \emph{Difmap} by shifting the maps by about a hundred times the beam size, corresponding to an off-center rms noise. We note that the rms noise in Stokes Q and U maps and in different sub-bands are similar and provide an average of them for each data set in (3) in Table~\ref{errorInfo}. $\sigma_{\Delta}$ is the D-term scatter (provided in (4) in Table~\ref{errorInfo}, see below), $N_{\rm ant}$ the number of antennas, $N_{\rm IF}$ the number of IFs, $N_{\rm scan}$ the number of scans with independent parallactic angles, and $I_{\rm peak}$ the peak of the total intensity map. We present the stations participating in the observations in Table~\ref{errorInfo}. In our case, $N_{\rm IF} = 1$ because we analyzed each sub-band data separately. We assumed $N_{\rm scan} = 8$ because all our data sets observed M87 as a primary target in a full-track observing mode. We assumed the error from imperfect EVPA calibration of $3^{\circ}$ because relatively small errors are expected, as can be seen in Figure~\ref{calRM} (see also Section~\ref{reduction}), and added this error to Equation~\ref{evpaerr} in quadrature.

\begin{deluxetable}{ccccccc}[t!]

\tablecaption{Information about data and errors}
\tablehead{
\colhead{Project code} & \colhead{Stations} &
\colhead{rms error [mJy/beam]} & \colhead{D-term scatter [\%]} & \colhead{$P_{\rm false}$} & \colhead{$N_{\rm false}$} & \colhead{$N_{\rm obs}$} \\
(1) & (2) & (3) & (4) & (5) & (6) & (7)
}
\startdata
 BJ020A & VLBA & 0.139 & 0.26 & $8.72\times10^{-5}$ & 38 & 3081 \\
 BJ020B & VLBA & 0.128 & 0.25 & $1.34\times10^{-6}$ & 0 & 2824 \\
 BC210B & VLBA, $-$MK, $-$OV & 0.090 & $0.41^*$ & $<3.64\times10^{-7}$ & 0 & 3427 \\
 BC210C & VLBA, $-$MK, $-$KP & 0.092 & $0.41^*$ & $<3.96\times10^{-7}$ & 0 & 4947 \\
 BC210D & VLBA, $-$KP, $-\frac{1}{2}$PT & 0.074 & 0.41 & $2.04\times10^{-5}$ & 7 & 1345\\
 BH135F & VLBA & 0.174 & $0.41^*$ & $2.50\times10^{-7}$ & 0 & 1125 \\
 BC167C & VLBA & 0.173 & $0.41^*$ & $<2.62\times10^{-7}$ & 0 & 863 \\
 BC167E & VLBA & 0.175 & $0.41^*$ & $1.33\times10^{-5}$ & 2 & 775
\enddata
\tablecomments{(1) Project code of VLBA observations. (2) VLBA stations participating in the observations. (3) Averages of off-center rms errors in Stokes Q and U maps in units of mJy/beam. (4) Scatter in the D-terms obtained with different sources in units of \%. We could not derive reliable D-term scatters in some data sets (marked with *) and thus assumed that the errors for these data sets are similar to that of session BC210D, $0.41\%$ (see Appendix~\ref{appendixa} for more details). (5) Probability of detecting false RMs with the RM values and $\chi_r^2$ similar to the observed ones (Appendix~\ref{appendixb}). $P_{\rm false}$ is obtained from integrating the FPDF between the minimum and maximum observed RMs for each data set presented in Table~\ref{RMvalue}. $<$ in front of the values for some sessions means that we could not find any pixel of false RM and we provide an upper limit. (6) Number of pixels of false RMs expected to be seen in the jet. (7) Number of pixels of observed RMs in the jet. \label{errorInfo}}

\end{deluxetable}

\begin{figure*}[!t]
\begin{center}
\includegraphics[trim=2mm 8mm 8mm 7mm, clip, width = 80mm]{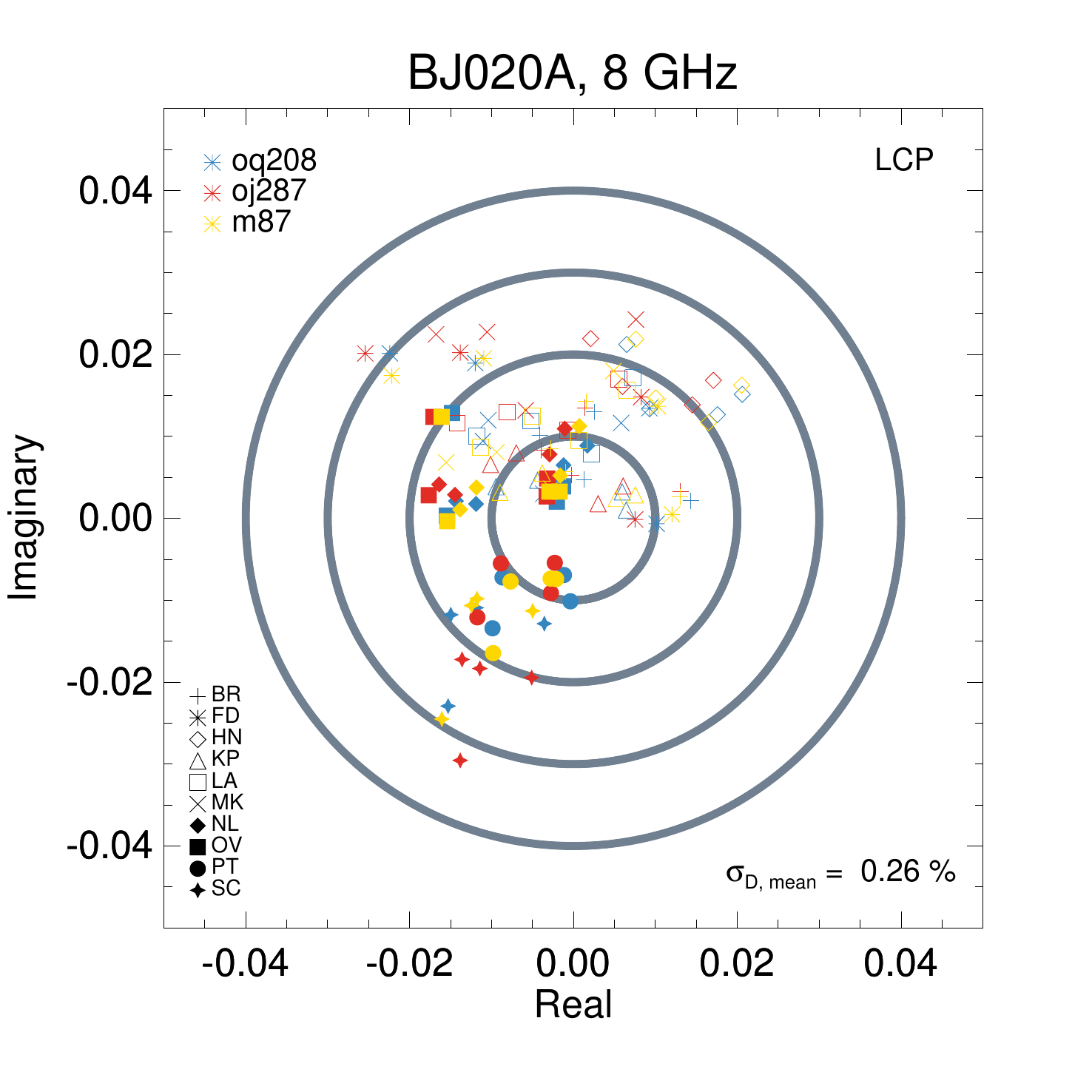}
\includegraphics[trim=2mm 8mm 8mm 7mm, clip, width = 80mm]{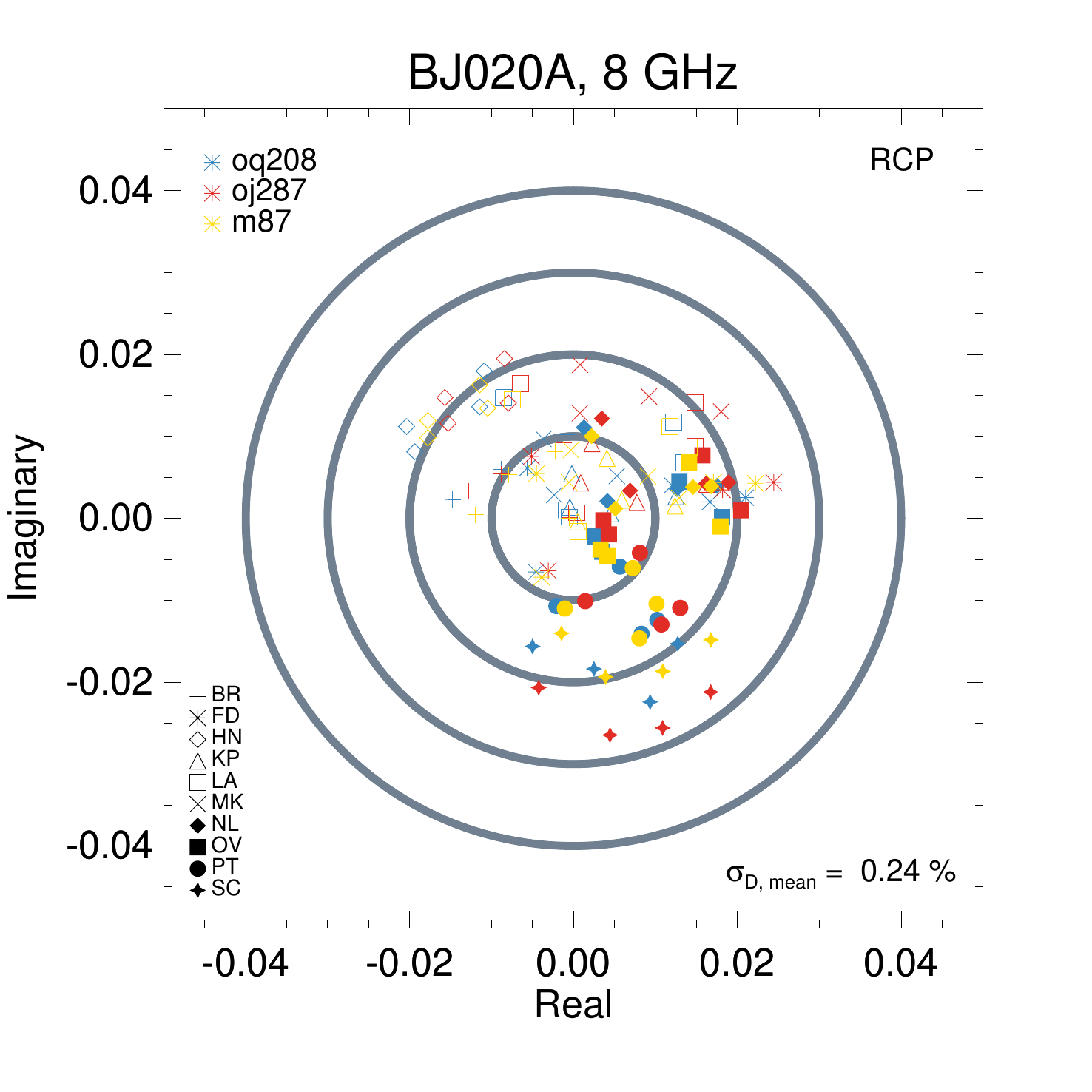}
\includegraphics[trim=2mm 8mm 8mm 7mm, clip, width = 80mm]{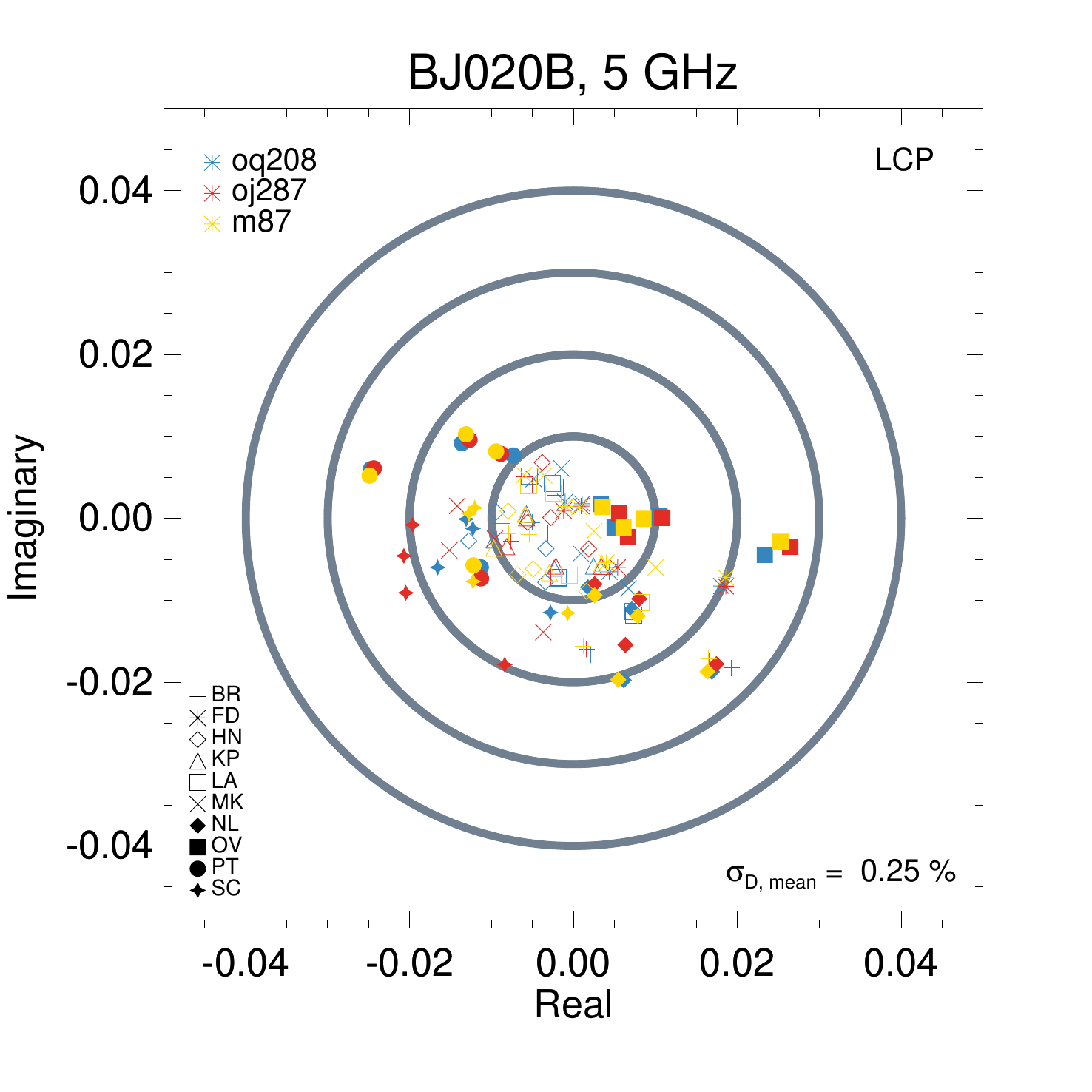}
\includegraphics[trim=2mm 8mm 8mm 7mm, clip, width = 80mm]{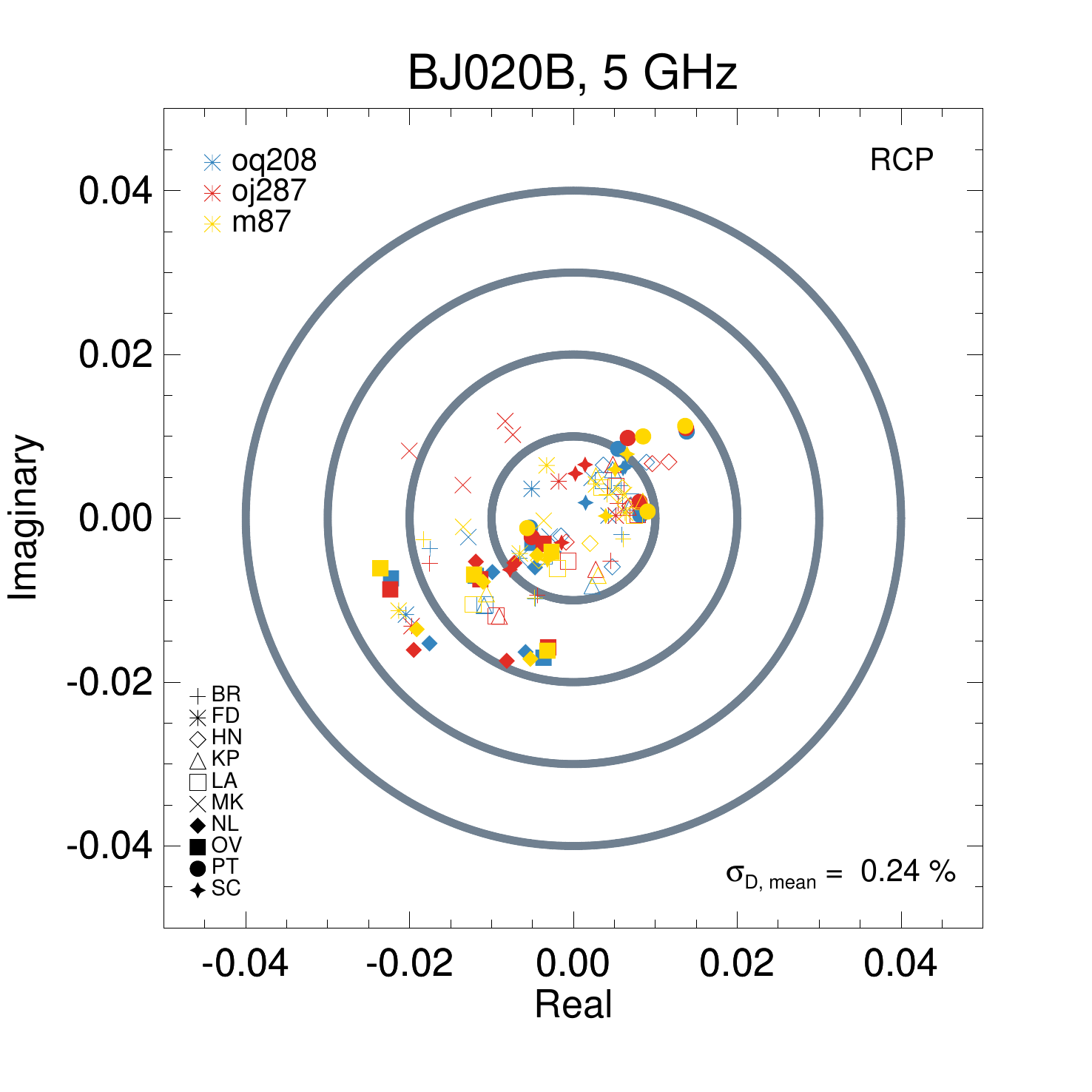}
\includegraphics[trim=2mm 8mm 8mm 7mm, clip, width = 80mm]{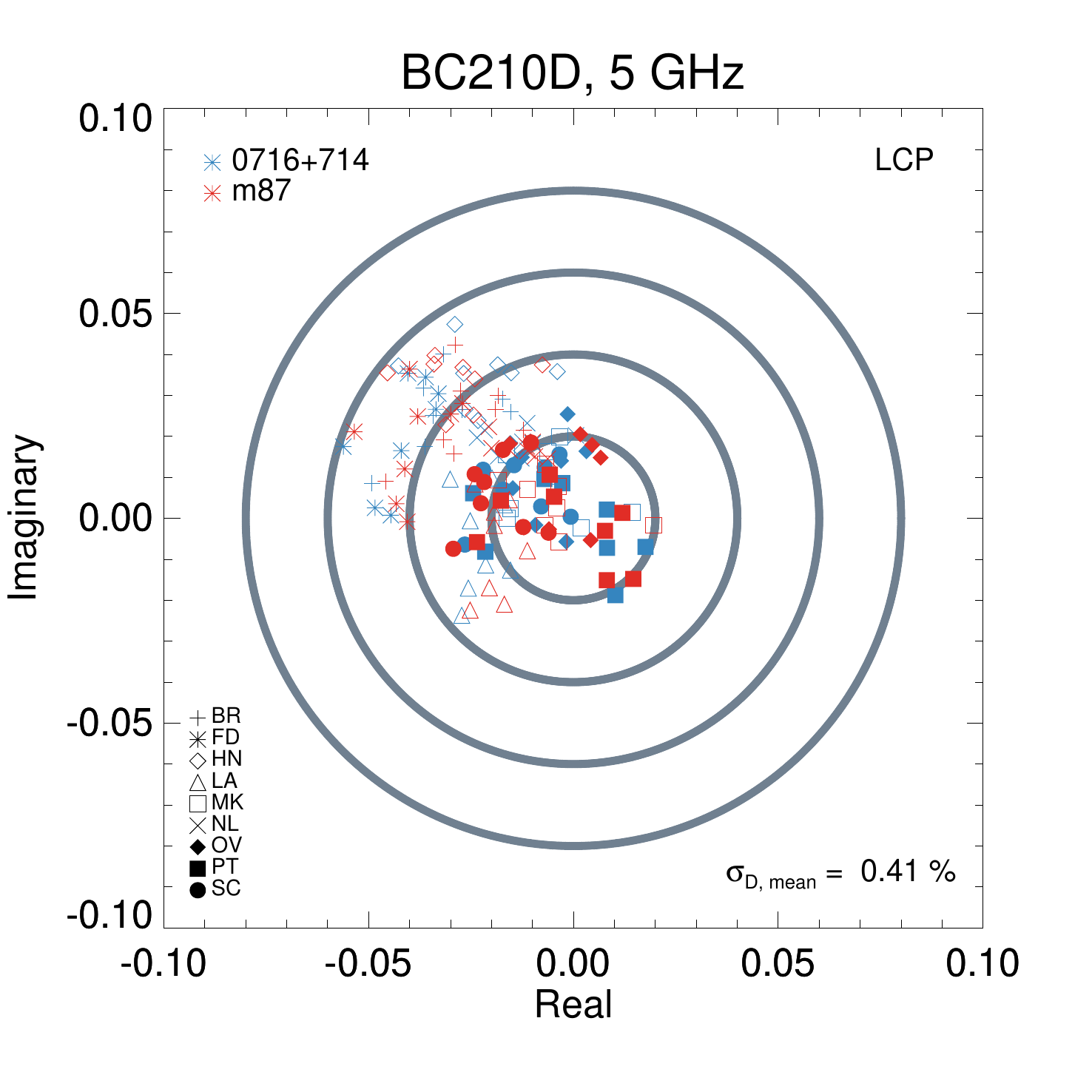}
\includegraphics[trim=2mm 8mm 8mm 7mm, clip, width = 80mm]{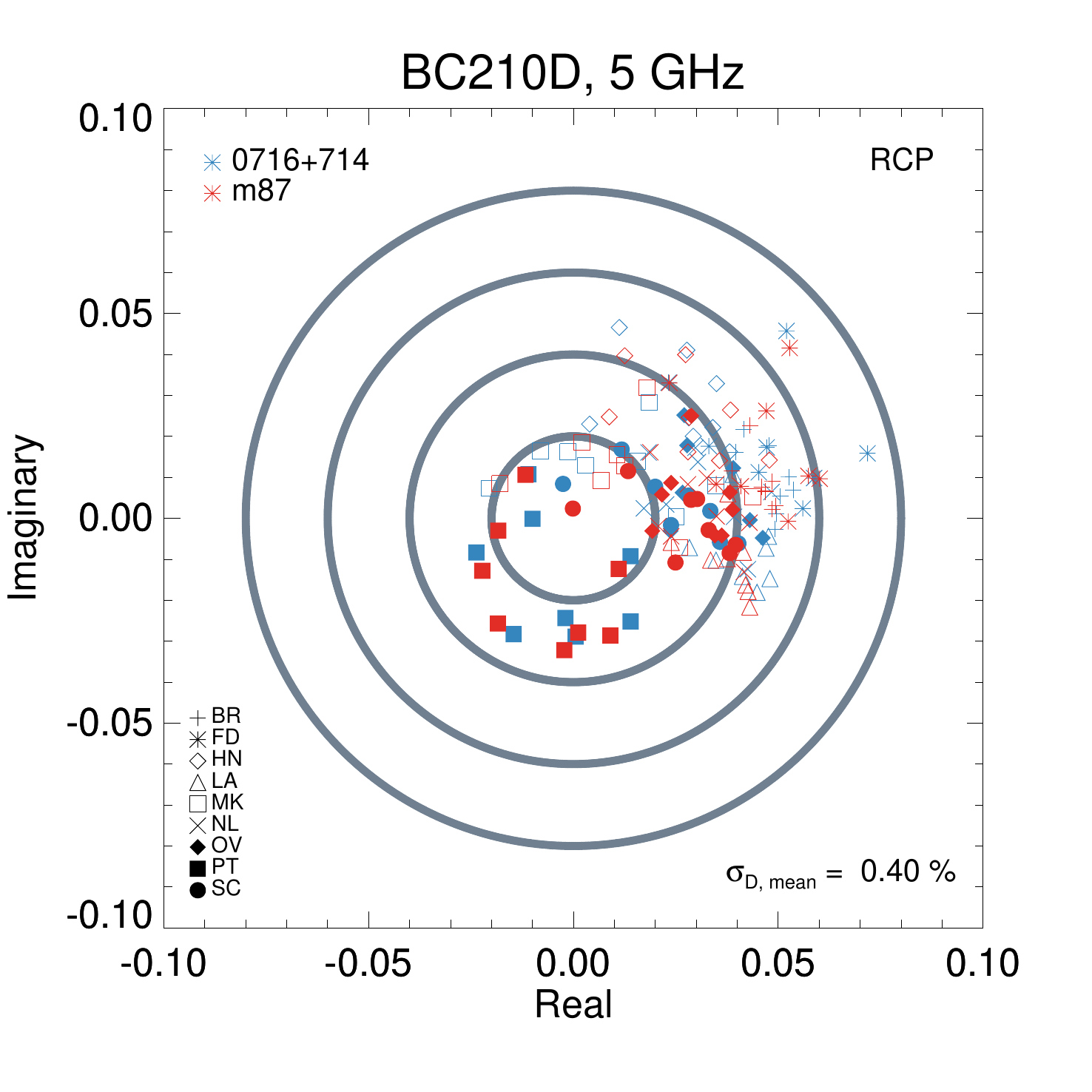}
\caption{D-terms obtained by using different calibrator sources (different colors) in the complex plane for the BJ020A, BJ020B, and BC210D data in the top, middle, and bottom panels, respectively. The left (right) panels are for the LCP (RCP) data. The average scatter in the D-terms is noted on the bottom right of each panel. \label{dterm}}
\end{center}
\end{figure*}

%

We estimated the D-term scatters by comparing the D-terms obtained from different sources. However, this was not always possible because some data sets did not have more than one source that is suitable for D-term calibration or because of a small number of scans (less than three) on other D-term calibrators. Specifically, we obtained reliable D-terms for BJ020A and BJ020B using three sources, OQ 208, OJ 287, and M87, because all of these are suitable for D-term calibration, i.e., either weakly polarized or moderately polarized but having compact geometries, and they are observed in multiple scans over large parallactic angle ranges (see e.g., \citealt{Roberts1994, Aaron1997, Park2018} for details of D-term calibration). We present the D-terms of different antennas obtained from different sources in Figure~\ref{dterm} (top for BJ020A and middle for BJ020B). The scatter in the D-terms obtained by using different sources is about $\approx0.25\%$ for both left-handed circularly polarized (LCP) and RCP data. 

For the BC210 data sets, we could obtain a reasonably small scatter of $\approx0.4\%$ only for the BC210D data (the bottom panel of Figure~\ref{dterm}), using M87 and 0716+714, because the number of scans on 0716+714 is at most two or three and two antennas were missing in the other two data sets. However, the D-terms measured by using M87 for BC210B and BC210C data sets are likely quite reliable because we obtained clear linear polarization in all different sub-bands that are consistent with the results of BJ020B and BC210D (Figure~\ref{stack}). Thus, we assumed that the D-term scatters for these data sets are similar to that of BC210D and used 0.41\%.

For the L band (2 GHz) data sets, we could not obtain the D-term scatters because only three sources, M87, 3C 273, and 3C 286, were observed. M87 can serve as a good D-term calibrator thanks to its very low degree of linear polarization. However, the other two sources show quite strong ($\gtrsim10\%$) linear polarization over large extended jet regions and thus they are not suitable for D-term calibration. We assumed that the D-term scatters of these data sets are the same as those of BC210D, i.e., 0.41\%. This is because the D-term scatters of the VLBA tend to be larger at higher observing frequencies (e.g., \citealt{Gomez2002}).


\section{Significance level of RM}
\label{appendixb}

We obtained RM for each pixel where the linear polarization intensity exceeds $1.5\sigma$ in all sub-bands, with $\sigma$ being the full uncertainty (Equation~\ref{polerr}, ~\ref{fullerr}). As there are at least four independently processed sub-bands per data set, the total (Gaussian) probability of false detection of RM is $<3.2\times10^{-4}$. However, linear polarization intensity does not follow a Gaussian probability distribution for a small signal-to-noise ratio \citep{WK1974, Trippe2014}. Thus, we need to carefully check the potential chance of detection of artifacts in the observed RMs. This kind of test has been done by performing extensive simulations in previous studies (e.g., \citealt{Roberts1994, Hovatta2012, Algaba2013, Mahmud2013}). They generate simulated data sets with the known polarized intensity distributions, e.g., a uniform fractional polarization and EVPA across the source's total intensity structure, and add errors introduced by various effects discussed in Appendix~\ref{appendixa}. The significance level can be inferred from the number of simulated data sets where the input polarized model is distorted.

However, this approach might not apply to our study because the observed linear polarization is very patchy in all data sets possibly due to substantial depolarization (Section~\ref{sectinternal}). We present an alternative approach to infer the significance levels of the observed RMs. If the criterion of $1.5\sigma$ cutoff is not strict enough and this introduces many false RMs in the jet, then one should expect to see many similar RMs outside the jet region (where the total intensity emission of the jet is not significant) as well. This is because all the error sources, i.e., random errors, CLEAN errors, and D-term errors can be distributed across the entire map, not specific to the jet region. Although the D-term errors depend on the total intensity (Equation~\ref{dtermerr}), this intensity is usually smaller than $\approx30$ mJy/beam where significant RMs are observed in the jet. Thus, the second term in Equation~\ref{dtermerr} is always dominant, and it would be fair to compare the observed RMs in the jet with the false RMs outside the jet region generated by errors.

For the regions outside the jet, we computed the number of pixels that satisfy the following two conditions: (i) polarized intensity above $1.5\sigma$ is detected in all sub-bands and (ii) $\chi_r^2 \lesssim 1.1-1.5$ are obtained for the $\lambda^2$ fit to the EVPAs, similarly to the observed RMs (We note, however, that the results are not significantly changed when we did not consider $\chi_r^2$). This calculation was done by using the maps having similar fields-of-view to those of the jet to properly compare with the observed jet RMs and to avoid the bandwidth-smearing and the time-average smearing effects. We obtained histograms of false RMs and divided them by the total number of pixels outside the jet region in the maps, which can serve as the false-alarm probability distribution functions (FPDFs) of detecting RM. 



\begin{figure}[!t]
\begin{center}
\includegraphics[trim=7mm 15mm 0mm 3mm, clip, width = 0.5\textwidth]{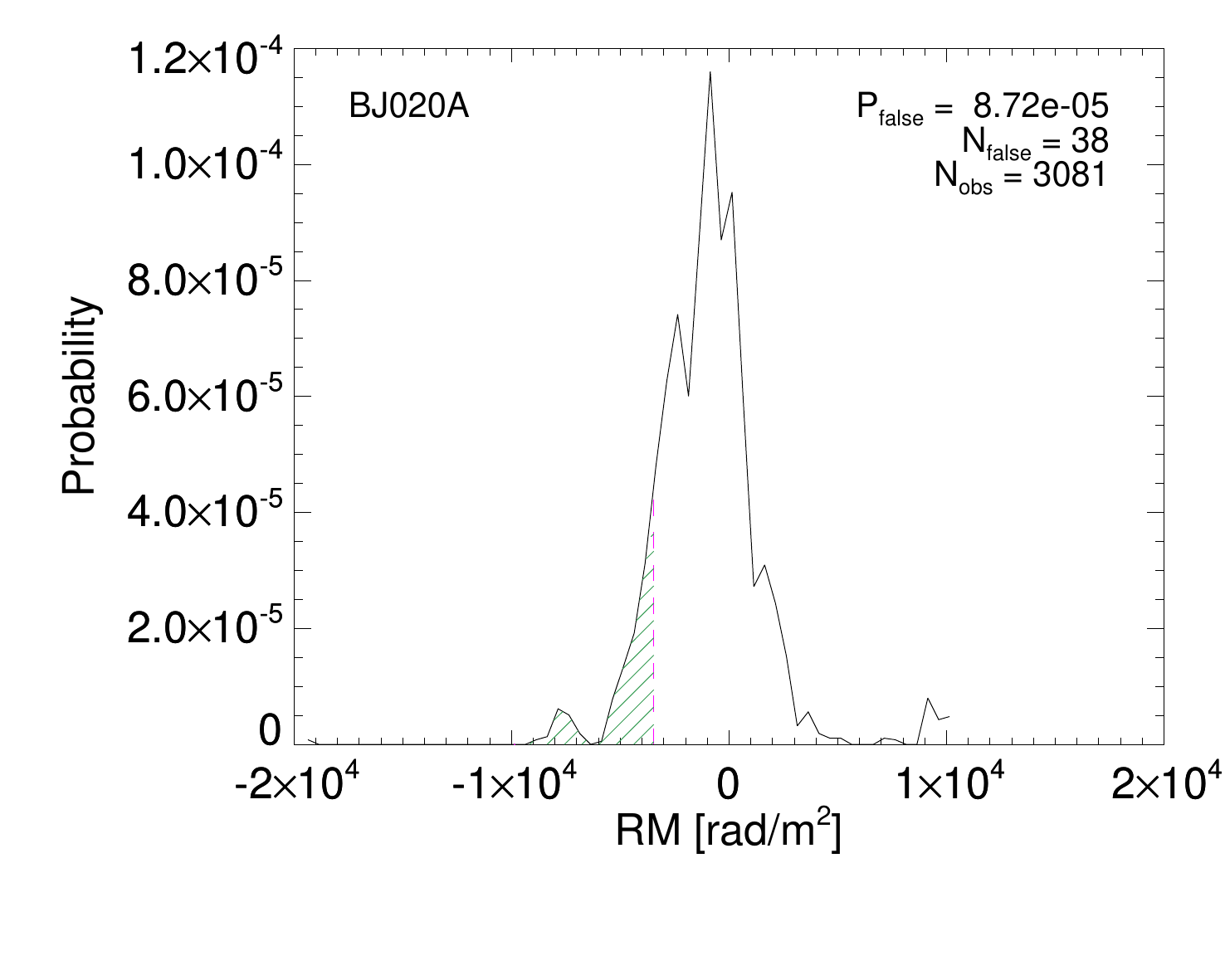}
\caption{False-alarm probability distribution function of detecting RM with the $1.5\sigma$ cutoff for BJ020A data set. $P_{\rm false}$ denotes a false-alarm probability of detecting RMs similar to the observed ones, obtained by integrating the green hatched region. $N_{\rm false}$ is the number of pixels of false RMs expected to be seen in the jet and $N_{\rm obs}$ the number of pixels of observed RMs in the jet. \label{false}}
\end{center}
\end{figure}

In Figure~\ref{false}, we present the FPDF for the 8 GHz data as an example. The probability of detecting false RMs with $-$10,163 $\lesssim \rm RM \lesssim$ $-$3,374 $\rm rad/m^2$ (the range of observed RMs at 8 GHz, see Table~\ref{RMvalue}) with good $\lambda^2$ fits, obtained from integrating the hatched region, is $8.72\times10^{-5}$ ($P_{\rm false}$). Accordingly, one can expect to detect false RMs in the jet region in approximately 38 pixels ($N_{\rm false}$), while significant RMs are detected in more than 3,000 pixels ($N_{\rm obs}$). We present the values of $P_{\rm false}$, $N_{\rm false}$, and $N_{\rm obs}$ for all data sets in Table~\ref{errorInfo}. The values of $P_{\rm false}$ for the other data sets, obtained by integrating the FPDF between the minimum and maximum observed RMs (Table~\ref{RMvalue}) for each data set, are even smaller, resulting in very small or zero $N_{\rm false}$.

In Table~\ref{Pfalse}, we present $P_{\rm false}$ obtained by using five different signal-to-noise ratio cutoffs. When  $1\sigma$ cutoff is used, $P_{\rm false}$ values are non-negligible, up to $\approx2\times10^{-3}$ for BJ020A data set. However, $P_{\rm false}$ decreases rapidly as the cutoff level increases for all data sets, becoming smaller than $\approx9\times10^{-5}$ with $1.5\sigma$ cutoff. Therefore, we conclude that almost all of the observed RMs obtained by using the $1.5\sigma$ cutoff is intrinsic to the source.

\section{RM maps for all observations}
\label{sectstack}

\begin{figure*}[!t]
\begin{center}
\includegraphics[trim=5mm 57mm 1mm 20mm, clip, width = \textwidth]{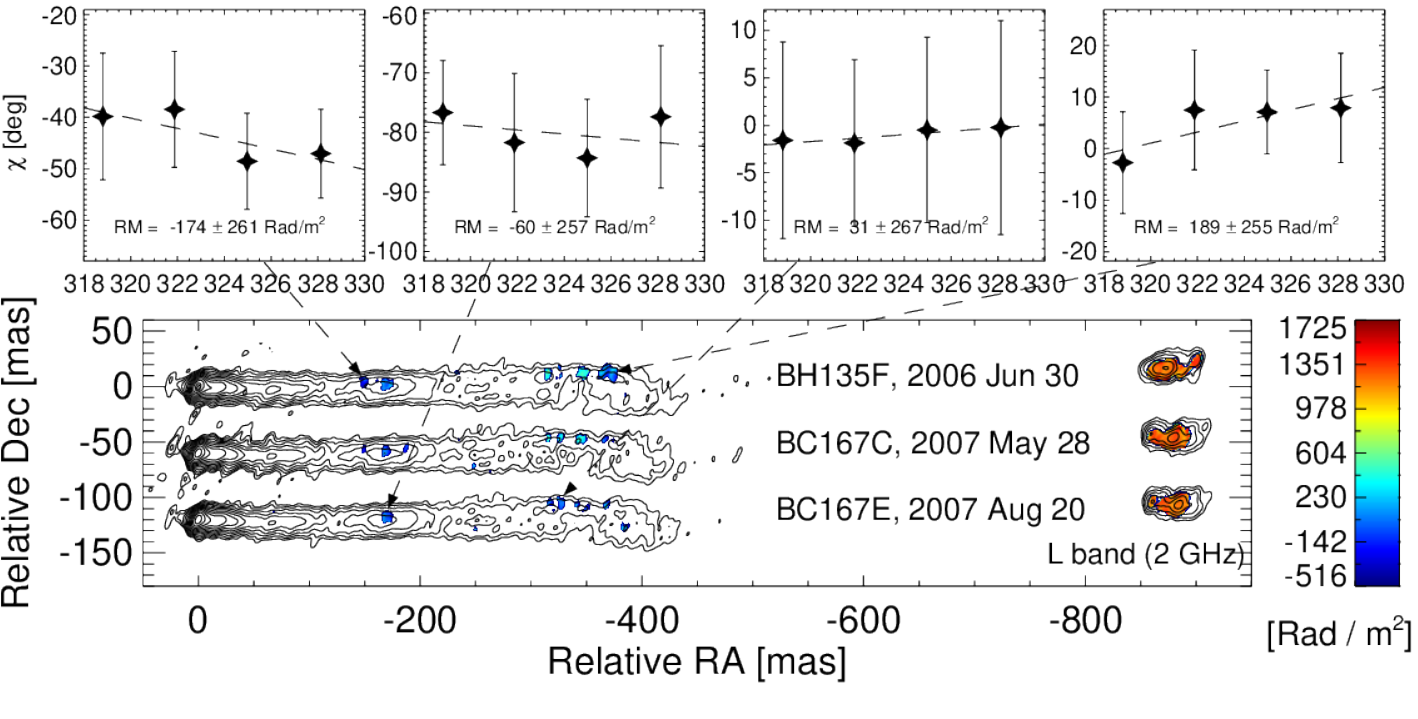}
\includegraphics[trim=7mm 60mm 1mm 6mm, clip, width = \textwidth]{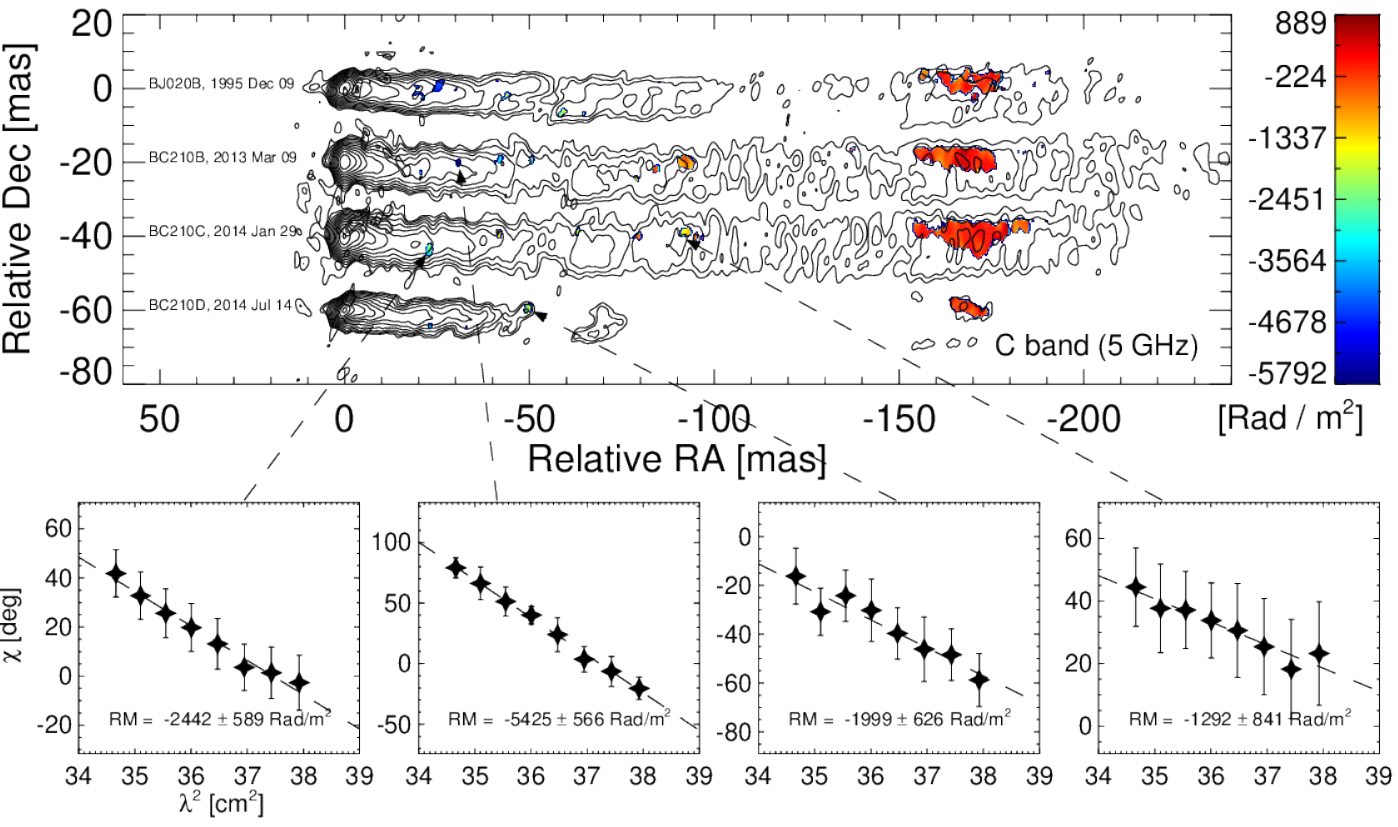}
\caption{RM maps and EVPA--$\lambda^2$ diagrams of three data sets we analyzed at 2 GHz (top) and four at 5 GHz (bottom). The maps in different epochs are shifted along the declination axis. All maps are rotated clockwise by $23^\circ$. The session and the observation date are noted for each map. Contours start at 0.79 and 0.60 mJy per beam for the 2 GHz and 5 GHz maps, respectively, and increase by factors of 2. Most of the extended jet emission is missing in the data from session BC210D because KP and half of PT antennas were missing (Table~\ref{errorInfo}), resulting in a significant loss of short baselines. \label{stack}}
\end{center}
\end{figure*}

We present the RM maps for the whole 2 and 5 GHz data sets in Figure~\ref{stack}. The RMs in different epochs at the same observing frequency are detected in similar locations of the jet, notably at $\approx170$ and $\approx320-370$ mas from the core at 2 GHz and at $\approx20-30$ and $\approx150-200$ mas from the core at 5 GHz. We note that the difference in the locations of some RMs might be due to relatively large time gaps (three months -- 17 years) between different data sets, given that the jet is known to move relativistically already at distances less than $\approx10$ mas from the core \citep{Mertens2016, Hada2017, Walker2018}.

\begin{deluxetable}{cccccc}[t!]
\tablecaption{$P_{\rm false}$ for different SNR cutoffs}
\tablehead{
 \multirow{2}{0.078\columnwidth}{Session} & \multicolumn{5}{c}{$P_{\rm false}$}\\
 & $1\sigma$ & $1.5\sigma$ & $2\sigma$ & $2.5\sigma$ & $3\sigma$
}
\startdata
 BJ020A & $2.45\times10^{-3}$ & $8.72\times10^{-5}$ & $8.00\times10^{-7}$ & $<2.67\times10^{-7}$ & $<2.67\times10^{-7}$ \\
 BJ020B & $4.32\times10^{-4}$ & $1.34\times10^{-6}$ & $<2.67\times10^{-7}$ & $<2.67\times10^{-7}$ & $<2.67\times10^{-7}$ \\
 BC210B & $5.89\times10^{-5}$ & $<3.64\times10^{-7}$ & $<3.64\times10^{-7}$ & $<3.64\times10^{-7}$ & $<3.64\times10^{-7}$ \\
 BC210C & $5.88\times10^{-4}$ & $<3.96\times10^{-7}$ & $<3.96\times10^{-7}$ & $<3.96\times10^{-7}$ & $<3.96\times10^{-7}$ \\
 BC210D & $2.43\times10^{-4}$ & $2.04\times10^{-5}$ & $<2.62\times10^{-7}$ & $<2.62\times10^{-7}$ & $<2.62\times10^{-7}$ \\
 BH135F & $2.66\times10^{-4}$ & $2.50\times10^{-7}$ & $<2.50\times10^{-7}$ & $<2.50\times10^{-7}$ & $<2.50\times10^{-7}$ \\
 BC167C & $2.15\times10^{-5}$ & $<2.62\times10^{-7}$ & $<2.62\times10^{-7}$ & $<2.62\times10^{-7}$ & $<2.62\times10^{-7}$ \\
 BC167E & $2.69\times10^{-4}$ & $9.06\times10^{-6}$ & $1.01\times10^{-6}$ & $<2.52\times10^{-7}$ & $<2.52\times10^{-7}$
\enddata
\tablecomments{$P_{\rm false}$ values using different signal-to-noise ratio cutoffs. $<$ in front of the values means that we could not find any pixel of false RM and we provide an upper limit. \label{Pfalse}}

\end{deluxetable}

\section{Radial RM profiles for the northern and southern jet edges}
\label{sectridge}

\begin{figure}[!t]
\begin{center}
\includegraphics[trim=22mm 18mm 10mm 3mm, clip, width = 88mm]{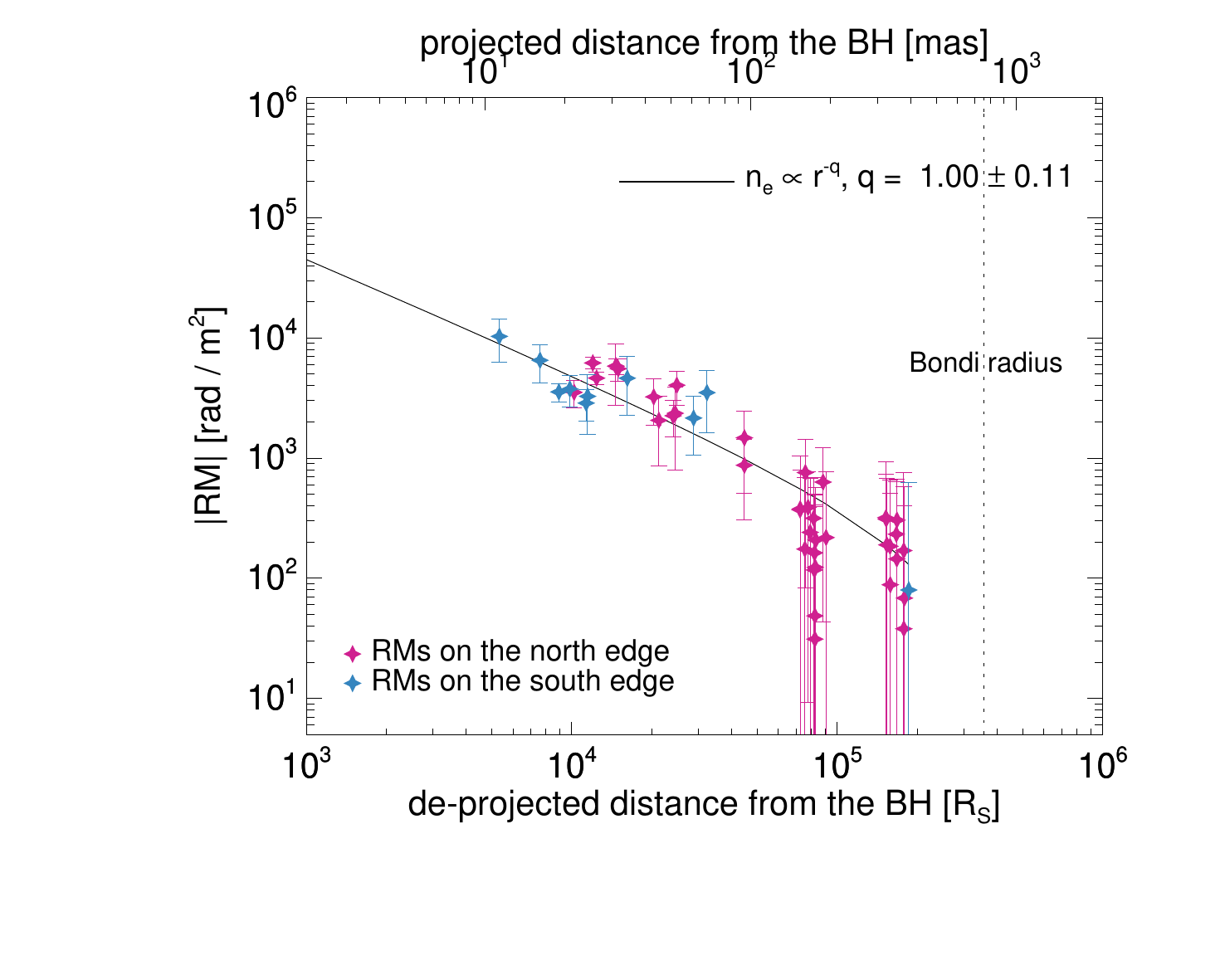}
\caption{Same as Figure~\ref{radial} but with data points obtained on the northern and southern jet edges shown in different colors.\label{ridge}}
\end{center}
\end{figure}

In Figure~\ref{ridge}, we present the absolute values of RM as a function of de-projected distance for the RMs detected on the northern and southern jet edges with different colors. We determined whether the observed RMs are located in the north or south edges by comparing the position of the RMs with the brightness centroid of the transverse intensity profile at the given distances. We found that out of 49 regions where significant RMs are detected, 10 are located in the southern jet edges.

\end{appendix}


\end{document}